\documentclass[cernpreprint, atlasdraft=true, texlive=2020, british, texmf]{atlasdoc}
 
\usepackage{atlaspackage}
\usepackage{atlasbiblatex}
 
\usepackage{atlasphysics}
 
\graphicspath{{logos/}{figures/}}
 
\usepackage{ANA-TOPQ-2018-06-PAPER-defs}

\AtlasTitle{Search for flavour-changing neutral-current interactions of a top quark and a gluon in $pp$ collisions at $\sqrt{s}=13$~TeV with the ATLAS detector}
 
\AtlasVersion{3.15}
\AtlasAbstract{
A search is presented for the production of a single top quark via left-handed flavour-changing neutral-current (FCNC) interactions of a top quark, a gluon and an up or charm quark. Two production processes are considered: $u+g\rightarrow t$ and $c+g\rightarrow t$. The analysis is based on proton--proton collision data taken at a centre-of-mass energy of 13$\,$TeV with the ATLAS detector at the LHC. The data set corresponds to an integrated luminosity of 139~fb$^{-1}$.
Events with exactly one electron or muon, exactly one $b$-tagged jet and missing transverse momentum are selected, resembling the decay products of a singly produced top quark. Neural networks based on kinematic variables differentiate between events from the two signal processes and events from background processes.
The measured data are consistent with the background-only hypothesis, and limits are set on the production cross-sections of the signal processes: $\sigma(u+g\rightarrow t)\times\mathcal{B}(t\rightarrow Wb)\times\mathcal{B}(W\rightarrow \ell\nu)<3.0\,$pb and $\sigma(c+g\rightarrow t)\times\mathcal{B}(t\rightarrow Wb)\times\mathcal{B}(W\rightarrow \ell\nu)<4.7\,$pb at the 95\% confidence level, with $\mathcal{B}(W\rightarrow \ell\nu)=0.325$ being the sum of branching ratios of all three leptonic decay modes of the $W$ boson.
Based on the framework of an effective field theory, the cross-section limits are translated into limits on the strengths of the $tug$ and $tcg$ couplings occurring in the theory: $|C^{\,ut}_{uG}|/\Lambda^2 < 0.057\,$TeV$^{-2}$ and $|C^{\,ct}_{uG}|/\Lambda^2 < 0.14\,$TeV$^{-2}$.
These bounds correspond to limits on the branching ratios of FCNC-induced top-quark decays: $\mathcal{B}(t\rightarrow u+g)< 0.61\times 10^{-4}$ and $\mathcal{B}(t\rightarrow c+g)< 3.7\times 10^{-4}$.}
 
\author{The ATLAS Collaboration}
 
\AtlasRefCode{TOPQ-2018-06}

\AtlasJournalRef{Eur. Phys. J. C 82 (2022) 334}
\PreprintIdNumber{CERN-EP-2021-234}
\AtlasDOI{10.1140/epjc/s10052-022-10182-7}

\hypersetup{pdftitle={ATLAS document},pdfauthor={The ATLAS Collaboration}}
 
\begin{document}
 
\maketitle
 
\tableofcontents

\section{Introduction}
\label{sec:intro}
 
Direct searches for on-shell production of new heavy particles at the
Large Hadron Collider (LHC) have not yet been successful. For this reason, indirect searches
targeting non-standard couplings among Standard Model (SM) particles attract
increasing interest. Among these analyses are searches for
flavour-changing neutral-current (FCNC) processes in the top-quark sector.
The SM does not contain FCNC processes at tree level, and even though these
processes exist at higher orders, they are suppressed due to the
Glashow--Iliopoulous--Maiani mechanism~\cite{Glashow:1970gm}.
Compared to the \bquark sector, where decays of $b$-hadrons via FCNCs were first observed in 1995~\cite{Alam:1994aw}, FCNC decays of top quarks are even more suppressed.
Depending on the decay mode, FCNC branching ratios ($\mathcal{B}$) of the top quark are predicted to
range from $10^{-12}$ to $10^{-17}$~\cite{AguilarSaavedra:2004wm}, and are thus well below the experimentally accessible regime, at present and in the foreseeable future.
The observation of FCNC top-quark decays or top-quark production via FCNCs would therefore be an
unambiguous signal of physics beyond the SM.
 
Many extensions of the SM predict significantly higher rates for FCNC processes in
the top-quark sector. These extensions include
new scalar particles introduced in two-Higgs-doublet models~\cite{Atwood:1996vj,Bejar:2000ub} or in
supersymmetry~\cite{Delepine:2004hr,Cao:2007dk,Yang:2018utw}.
In certain regions of the parameter space of these models, the predicted branching ratios of top quarks decaying via FCNC can be as large as $10^{-5}$ to $10^{-3}$
and thus become detectable at the LHC.
 
Searches for FCNCs involving a top quark and a gluon were performed at the Tevatron~\cite{Abazov:2010qk,Aaltonen:2008qr} and in data from Run~1 of the LHC~\cite{CMS_FCNC_8TeV,TOPQ-2014-13-witherratum,TOPQ-2011-18}.
Rather than looking for the top-quark decays $t\rightarrow u+g$ and $t\rightarrow c+g$ in top-quark--antiquark pair ($\ttbar$) production,
these analyses searched for the production of a single top quark ($t$) via the FCNC processes $u+g\rightarrow t$ ($ugt$ process) and $c+g\rightarrow t$ ($cgt$ process), exploiting specific kinematic features of single-top-quark production to separate a potential signal from the large $W+$jets and multijet backgrounds.
The analysis presented in this paper extends the Run~1 ATLAS search to the Run~2 data set collected with the ATLAS detector in the years 2015 to 2018, during which the
LHC operated at a centre-of-mass energy of $13\,\TeV$.
Conceptually, the scope of the analysis is expanded by performing independently optimised searches for the $ugt$ and $cgt$ processes.
Differences between these two processes are due to differences in the parton distribution functions (PDFs) for valence and sea quarks.
For top antiquarks the charge-conjugate processes are implied. The FCNC interaction is assumed to be left-handed.
Another novelty compared to the Run~1 analysis is the interpretation of the results in an effective field theory framework provided by the \texttt{TopFCNC} model~\cite{Degrande:2014tta}.
 
\begin{figure}[t]
\begin{center}
\includegraphics[width=0.35\textwidth]{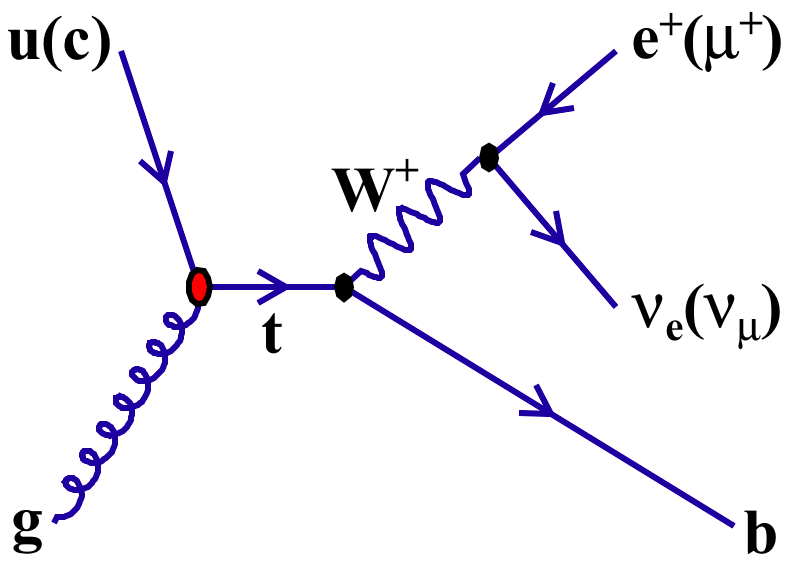}
\end{center}
\caption{Leading-order Feynman diagram of non-SM production of a single top quark via the FCNC
process $u(c)+g\rightarrow t$.}
\label{fig:qgt_Feynman}
\end{figure}

The event selection targets the $t\rightarrow e^+\nu b$ and
$t\rightarrow \mu^+\nu b$ decay modes of the top quark.
However, there is also additional but lower acceptance for events with the decay
$t\rightarrow \tau^+\nu b$ and the subsequent decay of the $\tau$-lepton into
$e^+\nu_e\bar{\nu}_\tau$ or $\mu^+\nu_\mu\bar{\nu}_\tau$.
A leading-order (LO) Feynman diagram illustrating the signature of the targeted
scattering events is shown in Figure~\ref{fig:qgt_Feynman}.
Considering the signature of the signal events,
the required reconstructed objects are exactly one charged-lepton
candidate (an electron or a muon) with high transverse momentum ($\pT$),
exactly one jet which is identified to originate with a high probability from a $b$-quark,
and large missing transverse momentum as an indication of a high-\pT neutrino.
 
The main background processes are $W{+}\,\bbbar$ production, $t$-channel single-top-quark ($tq$) production, $\ttbar$ production and multijet production.
Artificial neural networks (NNs) are used to separate signal events
from background events.
The observed distributions of the NN discriminants are analysed statistically with a
profile maximum-likelihood fit in which all systematic uncertainties are treated as
nuisance parameters.
 
The structure of the paper is as follows. A brief description of the ATLAS detector is given in Section~\ref{sec:detector}, followed by a comprehensive summary of the collision data and the samples of simulated events in Section~\ref{sec:samples}. Section~\ref{sec:selection} describes the reconstruction of detector-level objects and the event selection. The modelling of multijet background events and the estimation of their rate is discussed in Section~\ref{sec:multijet}. Section~\ref{sec:NN} provides details about the separation of signal and background events using NNs. Systematic uncertainties are outlined in Section~\ref{sec:systematics} and the results are presented in Section~\ref{sec:result}. Conclusions are given in Section~\ref{sec:conclusions}.
 
\section{The ATLAS detector}
\label{sec:detector}
\newcommand{\AtlasCoordFootnote}{
ATLAS uses a right-handed coordinate system with its origin at the nominal interaction point (IP)
in the centre of the detector and the \(z\)-axis along the beam pipe.
The \(x\)-axis points from the IP to the centre of the LHC ring,
and the \(y\)-axis points upwards.
Cylindrical coordinates \((r,\phi)\) are used in the transverse plane,
\(\phi\) being the azimuthal angle around the \(z\)-axis.
The pseudorapidity is defined in terms of the polar angle \(\theta\) as \(\eta = -\ln \tan(\theta/2)\).
Angular distance is measured in units of \(\Delta R \equiv \sqrt{(\Delta\eta)^{2} + (\Delta\phi)^{2}}\).}

The ATLAS detector~\cite{PERF-2007-01} at the LHC covers nearly the entire solid angle around the collision point.\footnote{\AtlasCoordFootnote}
It consists of an inner tracking detector surrounded by a thin superconducting solenoid, electromagnetic and hadronic calorimeters,
and a muon spectrometer incorporating three large superconducting toroidal magnets.
 
The inner-detector system (ID) is immersed in a \SI{2}{\tesla} axial magnetic field
and provides charged-particle tracking in the range \(|\eta| < 2.5\).
The high-granularity silicon pixel detector covers the vertex region and typically provides four measurements per track,
the first hit normally being in the insertable B-layer installed before Run~2~\cite{ATLAS-TDR-2010-19,PIX-2018-001}.
It is followed by the silicon microstrip tracker, which usually provides eight measurements per track.
These silicon detectors are complemented by the transition radiation tracker (TRT),
which enables radially extended track reconstruction up to \(|\eta| = 2.0\).
The TRT also provides electron identification information
based on the fraction of hits (typically 30 in total) above a higher energy-deposit threshold corresponding to transition radiation.
 
The calorimeter system covers the pseudorapidity range \(|\eta| < 4.9\).
Within the region \(|\eta|< 3.2\), electromagnetic calorimetry is provided by barrel and
endcap high-granularity lead/liquid-argon (LAr) calorimeters,
with an additional thin LAr presampler covering \(|\eta| < 1.8\)
to correct for energy loss in material upstream of the calorimeters.
Hadronic calorimetry is provided by the steel/scintillator-tile calorimeter,
segmented into three barrel structures within \(|\eta| < 1.7\), and two copper/LAr hadronic endcap calorimeters.
The solid angle coverage is completed with forward copper/LAr and tungsten/LAr calorimeter modules
optimised for electromagnetic and hadronic measurements respectively.
 
The muon spectrometer (MS) comprises separate trigger and
high-precision tracking chambers measuring the deflection of muons in a magnetic field generated by superconducting air-core toroids.
The field integral of the toroids ranges between \num{2.0} and \SI{6.0}{\tesla\metre}
across most of the detector.
A set of precision chambers covers the region \(|\eta| < 2.7\) with three layers of monitored drift tubes,
complemented by cathode-strip chambers in the forward region, where the background is highest.
The muon trigger system covers the range \(|\eta| < 2.4\) with resistive-plate chambers in the barrel, and thin-gap chambers in the endcap regions.
Interesting events are selected to be recorded by the first-level trigger system implemented in custom hardware,
followed by selections made by algorithms implemented in software in the high-level trigger~\cite{TRIG-2016-01}.
The first-level trigger accepts events from the \SI{40}{\MHz} bunch crossings at a rate below \SI{100}{\kHz},
which the high-level trigger reduces in order to record events to disk at about \SI{1}{\kHz}.
 
An extensive software suite~\cite{ATL-SOFT-PUB-2021-001} is used in the reconstruction and analysis of
real and simulated data, in detector operations, and in the trigger and data acquisition systems of the experiment.

 
\section{Samples of data and simulated events}
\label{sec:samples}
The analysis uses proton--proton ($pp$) collision data recorded with the ATLAS detector in the years 2015 to 2018 at a centre-of-mass energy of 13$\,\TeV$.
After applying data-quality requirements~\cite{DAPR-2018-01}, the data set corresponds to an integrated luminosity of $139\,\ifb$ with a relative uncertainty of $1.7\%$~\cite{ATLAS-CONF-2019-021}. The LUCID-2 detector~\cite{LUCID-2} was used for the primary luminosity measurements.
At the high instantaneous luminosity reached at the LHC, events were affected by additional inelastic $pp$ collisions in the same and neighbouring bunch crossings (pile-up). The average number of interactions per bunch crossing was 33.7.
 
Events were selected online during data taking by single-electron or single-muon triggers~\cite{TRIG-2018-05,Aad:2020uyd}.
Multiple triggers were used to increase the selection efficiency. The lowest-threshold triggers utilised isolation requirements for reducing the trigger rate. The isolated-lepton triggers had \pT thresholds of $20\,\GeV$ for muons and $24\,\GeV$ for electrons in 2015 data, and $26\,\GeV$ for both lepton types in 2016, 2017 and 2018 data. They were complemented by other triggers with higher \pT thresholds but no isolation requirements in order to increase the trigger efficiency.
 
Large sets of simulated events from signal and background processes were produced with event generator programs based on the Monte Carlo (MC) method to model the recorded and selected data.
After event generation, the response of the ATLAS detector was simulated using the $\GEANT$ toolkit~\cite{Agostinelli:2002hh} with a full detector model~\cite{SOFT-2010-01} or a fast simulation~\cite{ATL-PHYS-PUB-2010-013,AFII} which employed a parameterisation of the calorimeter response. To account for pile-up effects, minimum-bias interactions were superimposed on the hard-scattering events and the resulting events were weighted to reproduce the observed pile-up distribution. The minimum-bias events were simulated using \PYTHIA[8.186]~\cite{Sjostrand:2007gs} with the A3~\cite{ATL-PHYS-PUB-2016-017} set of tuned parameters and the \textsc{NNPDF2.3lo} PDF set~\cite{Ball:2012cx}.
Finally, the simulated events were reconstructed using the same software as applied to the collision data. Except for the multijet background, the same event selection requirements were applied and the selected events were passed through the same analysis chain. Small corrections were applied to simulated events such that the lepton trigger and reconstruction efficiencies, jet energy calibration and $b$-tagging efficiency were in better agreement with the response observed in data. More details of the simulated event samples are provided in the following subsections.
 
\subsection{Samples of simulated events from the $ugt$ and $cgt$ FCNC processes}
\label{sec:signal_samples}
Simulated events from the $ugt$ and $cgt$ processes were produced with the \METOP[1.0] event generator~\cite{Coimbra:2012ys,Guedes:2013voa} at next-to-leading order (NLO) in quantum chromodynamics (QCD). The difference between LO and NLO is very relevant for the analysis since a veto on a second jet is applied in the event selection by requiring exactly one reconstructed jet with $\pt>30\,\GeV$.
Signal samples generated at NLO predict a higher rate of events with two jets than samples generated at LO, leading to a lower acceptance for signal events due to the jet veto.
The Lorentz structure of the vertex coupling was taken to be left-handed.
It was verified that the shapes of kinematic distributions are independent of the value of the coupling constant used for the event generation.
The top quark was assumed to decay as in the SM and the decay was simulated using \MADSPIN~\cite{Frixione:2007zp,Artoisenet:2012st}.
Only leptonic decays of the $W$ boson originating from top-quark decay were considered, including $e^{\pm}$, $\mu^{\pm}$ and $\tau^{\pm}$ leptons. The renormalisation scale $\muR$ and the factorisation scale $\muF$ were set to the top-quark mass $m_t$, for which a value of $m_t=172.5\,\GeV$ was used. The CT10 set of PDFs~\cite{Gao:2013xoa} was used for event generation. Parton showers and the hadronisation were simulated with \PYTHIA[8.235]~\cite{Sjostrand:2014zea} with the A14 set of tune parameters~\cite{ATL-PHYS-PUB-2014-021}. In the {\METOP}+\PYTHIA set-up, hard gluon emissions can arise in both the NLO matrix-element generator and the parton-shower generator. The matching between the two generators was achieved by limiting the phase-space region of the first parton-shower emission in a way that depends on the transverse momentum of the top quark. The matching scale between the matrix-element generator and the parton shower was set to $10\,\GeV$.
 
Samples with alternative generator settings were produced to estimate systematic uncertainties. Samples with $\muR = \muF = 2 \cdot m_t$ and $\muR = \muF = 0.5 \cdot m_t$ were used to evaluate the impact of the scale choice on the signal model. The uncertainty in modelling parton showers was evaluated with \METOP signal samples in which parton showers were generated by \HERWIG[7.0.4]~\cite{Bahr:2008pv,Bellm:2015jjp} instead of \PYTHIA. The {\METOP}+{\HERWIG} set-up used the same PDF set as the nominal sample, CT10. In addition, {\METOP}+\PYTHIA samples with a different matching scale of $15\,\GeV$ were produced to evaluate the uncertainties due to the choice of this scale. All samples of the $ugt$ and $cgt$ processes were passed through the fast detector simulation.
 
\subsection{Simulation of \ttbar and SM single-top-quark production}
\label{sec:samples_top_quark}
Samples of simulated events from \ttbar and single-top-quark production were generated using the \POWHEGBOX[v2]~\cite{Nason:2004rx,Frixione:2007nw,Frixione:2007vw,Alioli:2009je,Alioli:2010xd,Re:2010bp,Frederix:2012dh} NLO matrix-element generator, setting $m_t=172.5\,\GeV$. For \ttbar and $tW$ production as well as $s$-channel single-top-quark production ($t\bar{b}$ production) the \textsc{NNPDF3.0nlo} PDF set~\cite{Ball:2014uwa} implementing the five-flavour scheme was used, while $t$-channel single-top-quark events ($tq$ production) were produced with the \textsc{NNPDF3.0nlo}\_nf4 PDF set, which implements the four-flavour scheme, following a recommendation given in Ref.~\cite{Frederix:2012dh}.
Parton showers, hadronisation, and the underlying event were modelled using \PYTHIA[8.230] with the A14 set of tuned parameters and the \textsc{NNPDF2.3lo} PDF set. The {\POWHEGBOX}+\PYTHIA generator set-up applies a matching scheme to the modelling of hard emissions in the two programs. The matrix-element-to-parton-shower matching is steered by the $h_\mathrm{damp}$ parameter, which controls the $\pt$ of the first additional gluon emission beyond the \LO Feynman diagram in the parton shower and therefore regulates the high-\pt emission against which the \ttbar system recoils. Event generation was run with $h_\mathrm{damp}=1.5\times m_t$~\cite{ATL-PHYS-PUB-2016-020}. The renormalisation and factorisation scales were set dynamically on an event-by-event basis, namely to $\muR = \muF = \sqrt{m_t^2 + \pt^2(t)}$ for \ttbar production and to $\muR = \muF = 4\,\sqrt{m_b^2 + \pt^2(b)}$ for $tq$ production, with $\pt(t)$ being the $\pt$ of the top quark and $\pt(b)$ being the $\pt$ of the $b$-quark originating from the initial-state gluon, splitting into a $\bbbar$ pair. The scale choice for $tq$ production followed a recommendation of Ref.~\cite{Frederix:2012dh}. When generating $tW$ events, the diagram-removal scheme~\cite{Frixione:2008yi} was employed to handle the interference with \ttbar\ production~\cite{ATL-PHYS-PUB-2016-020}.
 
In the case of $\ttbar$ production, top-quark decays were handled by \POWHEGBOX directly, while in the case of single-top-quark production, top-quark decays were modelled by \MADSPIN. The decays of bottom and charm hadrons were simulated using the \textsc{EvtGen~1.6.0} program~\cite{Lange:2001uf} for all samples involving top-quark production.
 
The \ttbar production cross-section was scaled to $\sigma(\ttbar)=832^{+47}_{-51}\,$pb, the  value obtained from next-to-next-to-leading-order (NNLO) predictions from the \textsc{Top++~2.0} program (see Ref.~\cite{Czakon:2011xx} and references therein), which includes the resummation of next-to-next-to-leading logarithmic (NNLL) soft-gluon terms. The total cross-sections for $tq$ and $t\bar{b}$ production were computed at NLO in QCD with the \textsc{Hathor~v2.1} program~\cite{Aliev:2010zk,Kant:2014oha} and the corresponding samples of simulated events were scaled to the following values: $\sigma(tq)=136.0^{+5.5}_{-4.7}\,$pb, $\sigma(\bar{t}q)=81.0^{+4.1}_{-3.7}\,$pb and $\sigma(t\bbar+\tbar b)=10.3\pm0.38\,$pb. The cross-section used for normalising the $tW$ sample is $\sigma(tW+\tbar W)=71.7\pm3.8\,$pb~\cite{Kidonakis:2010ux,Kidonakis:2013zqa}. All cross-section calculations assumed $m_t=172.5\,\GeV$ as a fixed value.
 
\subsection{Simulation of $W$+jets and $Z$+jets production}
The production of $W$ bosons and $Z$ bosons in association with jets, including heavy-flavour jets in particular, was simulated with the \SHERPA[2.2.1] generator~\cite{Bothmann:2019yzt}. In this set-up, NLO-accurate matrix elements for up to two partons and LO-accurate matrix elements for up to four partons were calculated with the \textsc{Comix}~\cite{Gleisberg:2008fv} and \OPENLOOPS[1]~\cite{Buccioni:2019sur,Cascioli:2011va,Denner:2016kdg} libraries. The default \SHERPA parton shower~\cite{Schumann:2007mg} based on Catani--Seymour dipole factorisation and the cluster hadronisation model~\cite{Winter:2003tt} were used. The generation employed the dedicated set of tuned parameters developed by the \SHERPA authors and the \textsc{NNPDF3.0nlo} PDF set.
 
The NLO matrix elements of a given jet multiplicity were matched to the parton shower using a colour-exact variant of the MC@NLO algorithm~\cite{Hoeche:2011fd}. Different jet multiplicities were then merged into an inclusive sample using an improved CKKW matching procedure~\cite{Catani:2001cc,Hoeche:2009rj} which was extended to NLO accuracy using the MEPS@NLO prescription~\cite{Hoeche:2012yf}. The merging threshold was set to $20\,\GeV$. The $W$+jets and $Z$+jets samples were normalised to NNLO predictions~\cite{Anastasiou:2003ds} of the total cross-sections, obtained with the \textsc{FEWZ} package~\cite{GAVIN20112388}.
 
\subsection{Simulation of diboson and multijet production}
\label{sec:sample_diboson_multijet}
Samples of on-shell diboson production ($WW$, $WZ$ and $ZZ$) were also simulated with the \SHERPA[2.2.1] generator. Motivated by the targeted signature of the signal events, only semileptonic final states were produced, in which one boson decayed leptonically and the other hadronically. The considered matrix elements contain all diagrams with four electroweak vertices and they were calculated at NLO accuracy in QCD for up to one additional parton and at LO accuracy for up to three additional parton emissions. The matching of NLO matrix elements to the parton shower and the merging of different jet multiplicities was done in the same way as for $W$/$Z$+jets production. Virtual QCD corrections were provided by the \OPENLOOPS[1] library. The \textsc{NNPDF3.0nlo} PDF set was used along with the dedicated set of tuned parameters developed by the \SHERPA authors. The diboson event samples were normalised to the total cross-sections provided by \SHERPA at NLO in QCD.
 
Events featuring generic high-$\pt$ multijet production may pass the event selection if a jet is misidentified as an electron or muon, or if real electrons or muons coming from hadron decays inside the jets pass the isolation requirements. The former are called \textit{fake leptons}, the latter \textit{non-prompt leptons}. In addition, non-prompt electrons occur as a result of photon conversions in the detector material.
Multijet events with fake electrons or non-prompt electrons were modelled with a sample of simulated dijet events, while events with non-prompt muons were modelled with collision data. The number of events with fake muons is negligible.
The dijet event sample was generated using \PYTHIA[8.186] with LO matrix elements for dijet production and interfaced to a \pt-ordered parton shower. The scales $\muR$ and $\muF$ were set to the square root of the geometric mean of the squared transverse masses of the two outgoing particles in the matrix element, $\muR = \muF = \sqrt[4]{(p_{\mathrm{T},1}^2 + m_1^2) (p_{\mathrm{T},2}^2 +  m_2^2)}$. At generator level, a filter was applied which required the existence of one particle-level jet with $\pt > 17\,\GeV$. The generation used the \textsc{NNPDF2.3lo} PDF set and the A14 set of tuned parameters. The generated sample of dijet events was used to model the event kinematics and to produce template distributions in the electron channel, while the rate of the multijet background was estimated in a data-driven way as described in Section~\ref{sec:multijet}.
 
\section{Object reconstruction and event selection}
\label{sec:selection}
\renewcommand\arraystretch{1.5}
The hard-scattering process was reconstructed by identifying the particles occurring at parton level with objects which were reconstructed at detector level, such as electron and muon candidates and hadronic jets. The presence of high-$\pt$ neutrinos is signalled by high missing transverse momentum.
 
\subsection{Object definitions}
Events were required to have at least one vertex reconstructed from at least two ID tracks with transverse momenta of $\pT>0.5\,\GeV$. The primary vertex of an event was defined as the vertex with the highest sum of $\pT^2$ over all associated ID tracks~\cite{ATL-PHYS-PUB-2015-026}.
 
Electron candidates were reconstructed from clusters of energy deposited in the electromagnetic calorimeter with a matched track reconstructed in the ID~\cite{EGAM-2018-01}. The pseudorapidity of clusters, $\eta_{\mathrm{cluster}}$, was required to be in the range $|\eta_{\mathrm{cluster}}|<2.47$. However, clusters were excluded if they are in the transition region $1.37 < |\eta_{\mathrm{cluster}}| < 1.52$ between the central and the endcap electromagnetic calorimeters. Electron candidates had to have $\pt>10\,\GeV$.
A likelihood-based method was used to simultaneously evaluate several properties of electron candidates, including shower shapes in the electromagnetic calorimeter, track quality, and detection of transition radiation produced in the TRT.
Two categories of electrons with different quality were defined~\cite{EGAM-2018-01}: the first category implemented \textit{Tight} identification criteria and featured a high rejection of non-prompt or fake electrons, while the second category with \textit{Loose} criteria had higher efficiency at the price of lower purity in prompt electrons.
Electrons from decays of weak gauge bosons pass the \textit{Tight} criteria with an average efficiency of 80\% and the \textit{Loose} criteria with 93\%.
 
Muon candidates were reconstructed by combining tracks in the MS with tracks in the ID~\cite{Aad:2020gmm}. The tracks had to be in the range of $|\eta|<2.5$ and have $\pt>10\,\GeV$. Similarly to electrons, two levels of identification criteria were applied, defining \textit{Medium} and \textit{Loose} quality categories of muon candidates. Muons orginating from $W$ bosons in $t\tbar$ events with $\pt>10\,\GeV$ pass the \textit{Medium} quality criteria with an efficiency of 97\% and the \textit{Loose} criteria with 99\%.
 
The tracks matched to electron and muon candidates had to point to the primary vertex, which was ensured by requirements imposed on the transverse impact-parameter significance, $|d_{0}/\sigma(d_{0})|<5.0$ for electrons and $|d_{0}/\sigma(d_{0})|<3.0$ for muons, and the longitudinal impact parameter, $|z_{0}\sin(\theta)|<0.5$~mm for both lepton flavours.
Isolated \textit{Tight} electrons and \textit{Medium} muons were selected by requiring the amount of energy in nearby energy depositions in the calorimeters and the scalar sum of the transverse momenta of nearby tracks in the ID to be small. Isolation requirements were not imposed on electrons and muons of \textit{Loose} quality.
Scale factors were used to correct the efficiencies in simulation in order to match the efficiencies measured for the electron~\cite{EGAM-2018-01} and muon~\cite{TRIG-2018-01} triggers, and the reconstruction, identification and isolation criteria.
 
Jets were reconstructed from topological clusters~\cite{ATL-PHYS-PUB-2015-036,PERF-2014-07} in the calorimeters with the anti-$k_t$ algorithm~\cite{Cacciari:2008gp} using FastJet~\cite{Cacciari:2011ma} and a radius parameter of 0.4. Their energy was calibrated~\cite{PERF-2016-04}, and they had to fulfil $\pt > 20~\GeV$ and $|\eta|<4.5$. Jets with $\pt < 120~\GeV$ and $|\eta|<2.5$ were required to pass a requirement on the jet-vertex-tagger (JVT) discriminant~\cite{PERF-2014-03} to suppress jets originating from pile-up collisions. The JVT-discriminant was required to be above $0.59$, which corresponds to an efficiency of $92\%$ for non-pile-up jets. Similarly, a forward-JVT (fJVT) requirement was used for jets with $\pt<60~\gev$ and $2.5<|\eta|<4.5$~\cite{PERF-2016-06}. Differences in the efficiencies of the JVT and fJVT requirements between collision data and simulation were accounted for by corresponding scale factors.
 
Jets containing $b$-hadrons were identified ($b$-tagged) with the \texttt{MV2c10} algorithm~\cite{FTAG-2018-01}, which used boosted decision tree discriminants with several $b$-tagging algorithms as inputs~\cite{PERF-2012-04}. The algorithms exploited the impact parameters of charged-particle tracks, the properties of reconstructed secondary vertices and the topology of $b$- and $c$-hadron decays inside the jets. In order to strongly reduce the misidentification rate of $c$-jets and light-flavour ($u$, $d$ or $s$) / gluon jets, a specific working point of the \texttt{MV2c10} algorithm was defined and calibrated, using the standard calibration technique~\cite{FTAG-2018-01}. With this working point, the $b$-tagging efficiency for jets that originate from the hadronisation of $b$-quarks is 30\% in simulated $\ttbar$ events. The $b$-tagging rejection\footnote{The rejection is defined as the inverse of the efficiency.} for jets that originate from the hadronisation of $c$-quarks ($u$-, $d$-, $s$-quarks or gluons) is $900$ (\num{30000}).
By using the high-purity $b$-tagging working point with 30\% efficiency for $b$-jets the analysis performance was considerably improved in comparison to an analysis based on the tightest standard working point which features a tagging efficiency of 60\% for $b$-jets. The improvement is mainly due to a reduced impact of the $W$+jets background, including uncertainties in mistagging $c$-quark jets, light-flavour jets and gluon jets in $W$+jets production.
Differences in $b$-tagging efficiency between simulated and collision events were corrected for by applying a $\pt$-dependent scale factor to simulated events. The scale factor ranges from $0.96\pm 0.04$ in the interval $30 < \pt(b) \leq 40\,\GeV$ to $1.01\pm 0.02$ for $140<\pt(b)<175\,\GeV$, which is the highest calibration interval relevant for this analysis. The $b$-tagging scale factors were obtained by comparing samples of collision data strongly enriched in $\ttbar$ events with samples of simulated events generated by $\POWHEG+$\PYTHIA[8.230]. The obtained scale factors depend on the parton-shower generator used to produce the $\ttbar$ samples. When using samples with a different parton-shower generator, for example \SHERPA to model \wjets events, or when evaluating systematic uncertainties with a set-up based on \HERWIG, additional correction factors called MC-to-MC scale factors were applied.
 
To avoid double-counting objects satisfying more than one selection criterion, a procedure called \textit{overlap removal} was applied. Reconstructed objects defined with \textit{Loose} quality criteria were removed in the following order:
electrons sharing an ID track with a muon;
jets within $\Delta R = 0.2$ of an electron, thereby avoiding double-counting electron energy deposits as jets;
electrons within $\Delta R = 0.4$ of a remaining jet, for reducing the impact of non-prompt electrons;
jets within $\Delta R = 0.2$ of a muon if they have two or fewer associated tracks;
muons within $\Delta R = 0.4$ of a remaining jet, reducing the rate of non-prompt muons.
The \textit{Tight} and \textit{Medium} criteria were applied to those objects which survived overlap removal.
 
The missing transverse momentum \vecpmiss was reconstructed as the negative vector sum of the $\pt$ of the reconstructed leptons and jets, as well as ID tracks that pointed to the primary vertex but were not associated with a reconstructed object~\cite{PERF-2016-07}. The magnitude of \vecpmiss is denoted by \met.
 
\subsection{Basic event selection}
\label{sec:basic_selection}
To be selected, events were required to have exactly one electron of \textit{Tight} quality or exactly one muon of \textit{Medium} quality, both with $\pt>27\,\GeV$. The charged lepton was required to match the object which triggered the event. To reduce contributions from $\ttbar$ events in the dilepton decay channel, any event with an additional lepton satisfying the \textit{Loose} quality conditions with $\pT > \SI{10}{\gev}$ was rejected (dilepton veto).
 
Multijet events containing fake or non-prompt leptons tend to have, in contrast to events with prompt leptons from $W$ and $Z$ decays, low \met and low $W$ transverse mass, which is defined as
\begin{equation}
\label{equ:mtw}
\mtw = \sqrt{2 \pT(\ell) \MET \left(1-\cos \Delta \phi \left(\ell, \vecpmiss \right) \right)}\,.
\end{equation}
To reduce the multijet background, $\met > 30\,\GeV$ and $\mtw > \SI{50}{\gev}$ were applied as additional selection requirements.
 
At least one jet with $\pt>30\,\GeV$ was required. In order to even further suppress the multijet background and to remove poorly reconstructed leptons with low $\pt$, the event selection applied an additional requirement based on the azimuthal angle between the primary lepton ($\ell$) and the leading jet ($j_1$), i.e.\ the jet with the largest $\pt$. This quantity is denoted by $\Delta \phi \left(j_1,\ell \right)$. The imposed requirement was
\begin{equation}\nonumber
\pT\left(\ell\right) > \SI{50}{\gev} \cdot \left( 1-\frac{\pi - |\Delta \phi\left(j_1, \ell \right)|}{\pi-1} \right)\,,
\end{equation}
which led to a tighter \pT requirement on the charged lepton if the leading jet and the lepton had a back-to-back topology, namely if $|\Delta \phi(j_1,\ell)|>0.687\pi$. For the maximum separation $|\Delta \phi(j_1,\ell)|=\pi$ between the two objects,
$\pT(\ell)>\SI{50}{\gev}$ had to be satisfied.
 
\subsection{Definition of signal and validation regions}
A signal region (SR) and three validation regions (VRs) were defined by applying further requirements to the sample of events passing the basic selection. Only events in the SR were used at a later stage of the analysis for a profile-likelihood fit to the data in the search for a signal contribution, while the VRs were used to validate the modelling of different background contributions.
A summary of the selection requirements used to define the four analysis regions is given in Table~\ref{tab:regions}.
\begin{table}[t]
\begin{center}
\caption{\label{tab:regions}Summary of selection requirements used to define the four analysis regions. The left column lists the observables on which the requirements are based. The first part of the table lists requirements which are common to all four analysis regions and define the basic event selection described in Section~\ref{sec:basic_selection}. Tight electrons and medium muons were counted based on a \pT threshold of $27\,\GeV$ and they are a subset of the corresponding Loose quality category. Loose charged leptons had to exceed a threshold of $\pT(\ell)=10\,\GeV$. The transverse mass of the $W$ boson, $\mtw$, is defined in Equation~(\ref{equ:mtw}). The efficiency of the $b$-tagging working point used to identify $b$-jets is denoted by $\epsilon_b$. The symbol $D_1$ represents one of the NN discriminants defined in Section~\ref{sec:NN}.}
\vspace*{3mm}
 
\begin{tabular}{ccccc}
\toprule
Observable & \multicolumn{4}{c}{Common requirements} \\ \midrule
$n_\mathrm{Tight}(e) + n_\mathrm{Medium}(\mu)$  & \multicolumn{4}{c}{$= 1$} \\
$n_\mathrm{Loose}(e) + n_\mathrm{Loose}(\mu)$  & \multicolumn{4}{c}{$= 1$} \\
$\met$    & \multicolumn{4}{c}{$> 30\,\GeV$} \\
$\mtw$  & \multicolumn{4}{c}{$ > \SI{50}{\gev}$} \\
$n(j)$  & \multicolumn{4}{c}{$\geq 1$} \\
$\pT\left(\ell\right)$ & \multicolumn{4}{c}{$> \SI{50}{\gev} \cdot \left( 1-\frac{\pi - |\Delta \phi\left(j_1, \ell \right)|}{\pi-1} \right)$}\\ \midrule
& \multicolumn{4}{c}{Analysis regions} \\
& SR & \wvr & \ttvr & \tchanvr \\ \midrule
$n(|\eta(j)|<2.5)$ & $=1$ & $=1$ & $=2$ & $=1$ \\
$n(b)$  &  $=1$ & $=1$ & $=2$ & $=1$ \\
$\epsilon_b$ & 30\% & 60\% (veto 30\%) & 30\% & 30\% \\
$n(|\eta(j)|>2.5)$ & $\geq 0$ & $\geq 0$ & $\geq 0$ & $=1$ \\
$D_{1(2)}$   & -- & $0.3 < D_{1(2)} < 0.6 $ & -- & $0.2 < D_{1(2)} < 0.4 $ \\ \bottomrule
\end{tabular}
\end{center}
\end{table}
All requirements mentioned before are common to all regions considered. The SR was defined by narrowing the jet requirement relative to the basic event selection. Each event had to have exactly one jet with $\pt>30\,\GeV$ and $|\eta|<2.5$, i.e.\ events with additional central jets were vetoed. This single jet had to be $b$-tagged. The selection efficiency for signal events in which the top quark decays into $Wb$ and the resulting $W$ boson decays leptonically was $1.36$\% for $ugt$ events and $2.30$\% for $cgt$ events.
For the $ugt$ search, the SR was split according to the sign of the charge of the primary lepton $\sgn q(\ell)$. Two NN discriminants $D_1$ and $D_2$, described in Section~\ref{sec:NN}, were formed to separate signal and background events in these three SRs.
 
The first VR was defined for validating the modelling of the events kinematics of \wjets production (\wvr) by the \SHERPA[2.2.1] generator. To suppress top-quark backgrounds a less stringent $b$-tagging requirement was used.
Exactly one jet with $\pt>30\,\GeV$ was required to be $b$-tagged at a working point with an efficiency of 60\%. All other selection requirements were the same as for the SR. However, events in the SR were vetoed.
The modified $b$-tagging requirement leads to a different flavour composition of the jets in the \wvr compared to the SR; the components of $W$+$c$-jets and $W+$light-flavour jets are increased relative to $W$+$b$-jets.
To enrich the region further in \wjets events and reduce the number of signal events, the NN discriminant $D_1$, specified in Section~\ref{sec:NN}, was required to be in the range $0.3 < D_1 < 0.6 $. The modelling of events with positive lepton charge was separately checked by requiring the NN discriminant $D_2$ to be in the range $0.3 < D_2 < 0.6 $, defining the $\ell^+\;$\wvr.
When normalising the FCNC processes to the observed limits from the previous ATLAS results obtained at a centre-of-mass energy of 8\,\TeV, the FCNC signal contamination is 1.2\% in the \wvr and 0.9\% in the $\ell^+\;$\wvr.
 
The second VR was enriched in \ttbar events by selecting events with exactly two $b$-tagged jets using the 30\% $b$-tagging working point (\ttvr). When normalising the FCNC processes to the observed limits from the previous ATLAS results obtained at a centre-of-mass energy of 8\,\TeV, the FCNC signal contamination is at a very low level of a few times $10^{-4}$.
The third VR checked the modelling of $tq$ events (\tchanvr). Events with exactly two jets were required. Exactly one of the jets had to be $b$-tagged at the 30\% efficiency working point, while the second jet was required to be in the forward region with $|\eta|>2.5$, which is a characteristic feature of $tq$ events. Thus, the \tchanvr was a subset of the SR, since there was no condition on jets in the forward region when defining the SR. To further enhance the fraction of $tq$ events and to suppress signal events, the NN discriminant $D_1$ was required to be in the range $0.2 < D_1 < 0.4$. The modelling of events with positive lepton charge was	separately checked by requiring	the NN discriminant $D_2$ to be in the range $0.2 < D_2 < 0.4 $, defining the $\ell^+\;$\tchanvr.
When normalising the FCNC processes to the observed limits from the previous ATLAS results obtained at a centre-of-mass energy of 8\,\TeV, the FCNC signal contamination is 1.2\% in the \tchanvr ($cgt$ analysis) and 0.8\% in the $\ell^+\;$\tchanvr.
 
 
\section{Estimation of the multijet background}
\label{sec:multijet}
 
By requiring electron and muon candidates to be isolated, the object definition and the event selection strongly favour prompt leptons originating from decays of $W$ bosons or $Z$ bosons. However, there is a small probability for non-prompt electrons or muons occurring in hadron decays, either directly or through the decay of a $\tau$-lepton, to be reconstructed as isolated leptons. The main source is $b$-hadron decays in jets, but $c$-hadrons and long-lived weakly decaying states such as $\pi^\pm$ and $K$ mesons also contribute. In addition, prompt electrons are mimicked by \textit{fake electrons} which arise from the misidentification of direct photons, photons from $\pi^0$ decays, or bremsstrahlung and photon conversions. Even though the probabilities of misidentification are relatively low, some multijet events still pass the selection and contribute to the background, since their production cross-section is approximately three orders of magnitude higher than the cross-sections of top-quark production processes. As the mechanisms of misidentification are not well modelled by the detector simulation, the rate of the multijet background was determined in a data-driven way by fitting the $\met$ distribution for events with an electron (electron channel) and the $\mtw$ distribution for events with a muon (muon channel).
 
In the electron channel, the multijet background was modelled using the \textit{jet-electron} method~\cite{ATLAS-CONF-2014-058}. Simulated events from dijet production (see Section~\ref{sec:sample_diboson_multijet} for a description of the sample) were selected if they contained a jet depositing a large fraction (${>}80\%$) of its energy in the electromagnetic calorimeter. This jet was classified as an electron, the jet-electron, and treated in the subsequent steps of the analysis in the same way as a properly identified prompt electron. The jet-electrons had to pass the nominal $\pt$ and $|\eta|$ requirements, but electron identification requirements were not applied. Since the relative numbers of electrons detected in the barrel ($|\eta|<1.37$) and endcap ($|\eta|>1.52$) sections of the electromagnetic calorimeter were not modelled well enough by the sample of simulated dijet events, the electron channel was divided into two subchannels: a barrel-electron channel and an endcap-electron channel.
 
In the muon channel, multijet events were modelled with collision events highly enriched in non-prompt muons~\cite{ATLAS-CONF-2014-058}. Starting from the same sample of collision events as the nominal selection, a subset of events enriched in non-prompt muons was obtained by inverting or modifying some of the muon isolation requirements, such that the resulting sample did not overlap with the nominal sample. The kinematic requirements on muon $\pt$ and $|\eta|$ remained the same as for the nominal selection.
 
The rate of the multijet background was normalised by performing a binned maximum-likelihood fit to the $\met$ and $\mtw$ distributions observed in the electron and muon channels, respectively. All selection criteria were applied, except for the $\met$ requirement in the electron channels (barrel and endcap) and the requirement on $\mtw$ in the muon channel. The three channels were further split according to the sign of the charge of the primary lepton $\sgn q(\ell)$, leading to six channels per analysis region. Separate fits were performed for the SR and the three VRs. In each region, all six channels were fit simultaneously. Since the multijet background is expected to be independent of lepton charge, its rates in the $\ell^+$ and the $\ell^-$ channels were assumed to be the same. On the other hand, the rates of some of the other background processes, i.e.\ $tq$, $t\bar{b}$ and \wjets production, are different in the $\ell^+$ and the $\ell^-$ channels due to the PDFs. For the purpose of these fits, scattering processes other than multijet production were grouped in the following way: (1) \textit{top-quark production} comprises $\ttbar$ production and all three single-top-quark production processes ($tq$, $t\bar{b}$ and $tW$ production), (2) \wjets production, including the production of light-quark, gluon, $b$-quark and $c$-quark jets in association with a $W$ boson, and (3) $Z$+jets and diboson production ($WW$, $WZ$ and $ZZ$ production). The templates of the fit distributions for these three groups of processes were derived from simulated events and the rates were normalised to the theory predictions reported in Section~\ref{sec:samples}. As the shapes of the distributions for $Z$+jets and diboson production are very similar to those of \wjets production, the rates of $Z$+jets and diboson production were fixed in the fitting process to the values predicted by simulation. Uncertainties in the normalisation of top-quark production and \wjets production were accounted for by Gaussian constraints on the normalisation factors of these groups of processes. In the \wvr, only the rate of $W$+jets production was varied, while the top-quark background was fixed. Similarly, in the \ttvr and \tchanvr only the rate of top-quark production was varied, while the rate of \wjets production was fixed. In the SR, both rates were free to vary within uncertainties.
 
The fits yielded estimates of the rates of the multijet background in the four analysis regions before applying the requirements on $\met$ and $\mtw$. An uncertainty of $30\%$ was assigned to the estimates, covering alternative results obtained in studies of fits to different discriminating observables. As examples illustrating the fit results, Figure~\ref{fig:illustration_QCDfit_SR} shows the $\met$ distribution in the $e^+$ barrel channel of the SR and the $\mtw$ distribution in the $\mu^+$ channel of the SR.
\begin{figure}[!t!b]
\centering
\subfloat[]{
\includegraphics[width=0.45\textwidth]{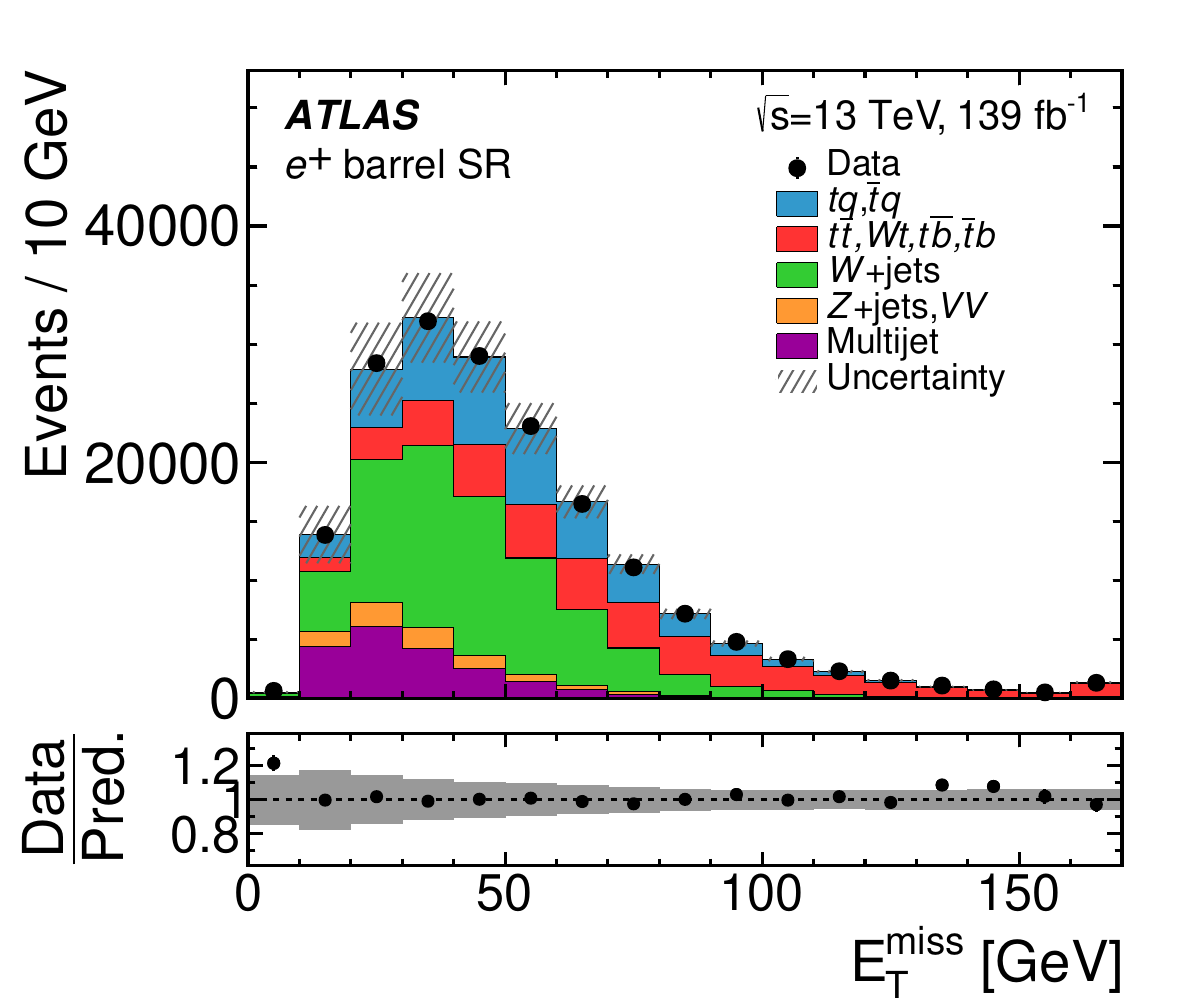}
\label{fig:barrel_ele_plus_QCDFit_SR}}\qquad
\subfloat[]{
\includegraphics[width=0.45\textwidth]{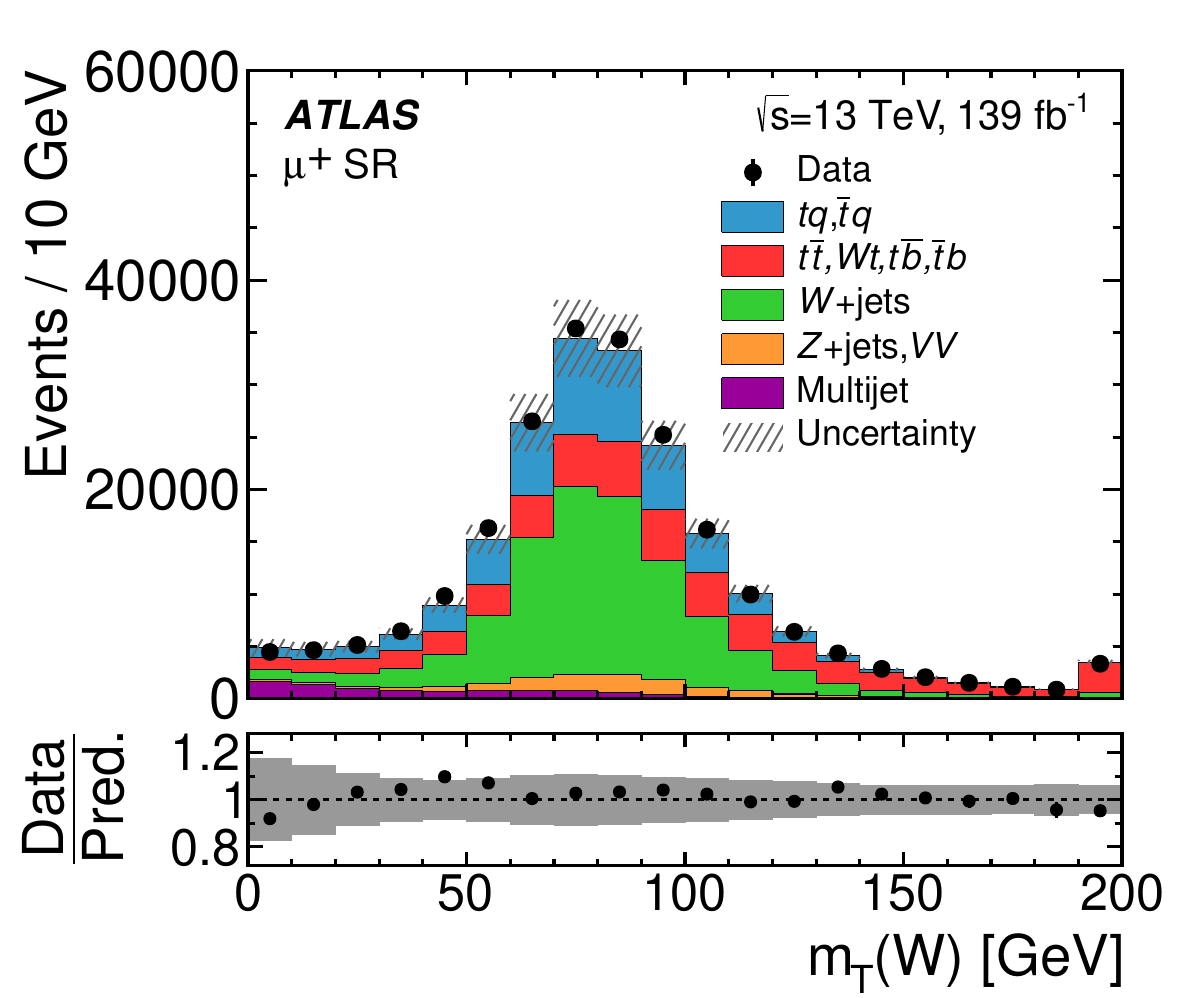}
\label{fig:muon_plus_QCDFit_SR}}
\caption[]{Illustration of the estimation of the multijet background by fitting the $\met$ and $\mtw$ distributions in the analysis regions. As representative examples, the $\met$ distribution is shown in the $e^+$ barrel channel in~\subref{fig:barrel_ele_plus_QCDFit_SR} and the $\mtw$ distribution is shown in the $\mu^+$ channel in~\subref{fig:muon_plus_QCDFit_SR}. Both distributions are in the SR. The stacked histograms were normalised to the fit result. The uncertainty band represents the uncertainty due to limited sample size and the rate uncertainties of the different processes ($20$\% for $W$+jets production, $30$\% for the multijet background and $6$\% for the top-quark processes). The ratio of observed to predicted (Pred.) numbers of events in each bin is shown in the lower panel. Events beyond the axis range are included in the last bin.}
\label{fig:illustration_QCDfit_SR}
\end{figure}
The stacked histograms were normalised to the fit result. The low $\met$ and $\mtw$ regions drove the estimate of the multijet background, since its fraction of the total yield was larger there than at higher values of the two observables. The yield of the multijet background after applying the requirements of $\met>30\,\gev$ and $\mtw>50\,\gev$ is based on the normalised histograms of the multijet background normalised to the fit result and was later used as a starting value for the profile-likelihood fit in the final statistical analysis. The normalisation factors obtained for top-quark production and \wjets production were applied to normalise the respective backgrounds when validating the modelling of kinematic distributions prior to the statistical analysis of the NN discriminants, but they were not used in the statistical analysis itself.
 
All backgrounds other than the multijet background were modelled by simulated events and the event rate was estimated by scaling the samples of simulated events to the integrated luminosity of the sample of collision data being analysed. The event kinematics of the multijet background is described with the jet-electron model and with non-prompt muon events, normalising the rate of the multijet background to the results of the fits to the $\met$ and $\mtw$ distributions. Figure~\ref{fig:pie_chart_SR} provides a summary of the fractional contributions of the different background processes to the expected event yield in the SR.
\begin{figure}[!t!b]
\centering
\includegraphics[width=0.55\textwidth]{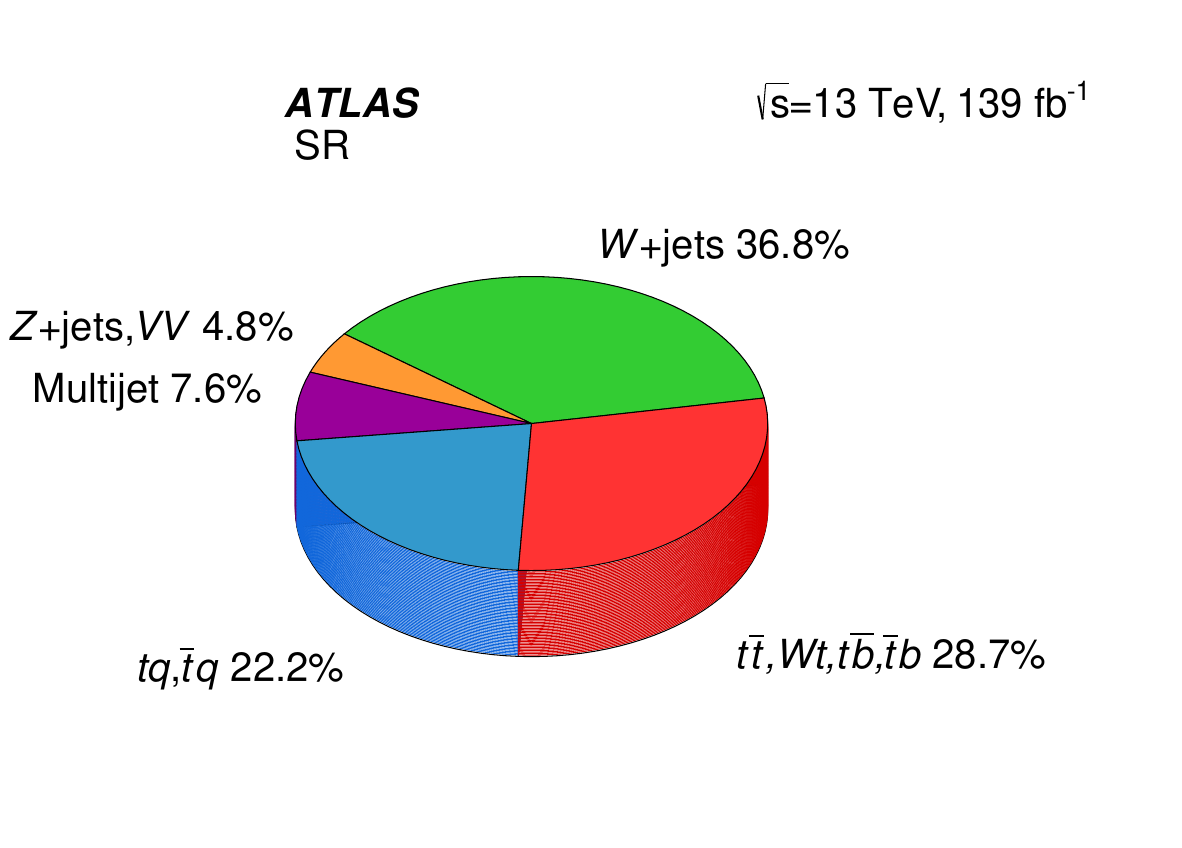}
\caption[]{\label{fig:pie_chart_SR}Pie chart of the background composition of the SR. The SR comprises the two electron channels (barrel and endcap) and the muon channel. The pre-fit event yields are reported in Table~\ref{tab:res_evnt_yield_final}.}
\end{figure}
The three largest backgrounds are \wjets production, the combined $t\tbar$-$tW$-$t\bbar$ background, and $tq$ production.
\FloatBarrier
 
\section{Neural networks separating signal and background events}
\label{sec:NN}
\renewcommand\arraystretch{1.3}
Two NNs were employed to enhance the separation of signal events from background events by combining several kinematic (input) variables to form two discriminants named $D_1$ and $D_2$. The kinematics of signal events depends on whether the quark (antiquark) in the initial state is a valence quark or a sea quark (antiquark).
Sea quarks (antiquarks) and valence quarks of the proton carry, on average, different fractions $x$ of the proton momentum and this difference leads to different rapidity distributions for the corresponding produced top quarks (antiquarks) and their decay products.
Top quarks produced in the $u+g\rightarrow t$ process tend to have higher absolute rapidity values than top antiquarks produced in the $\ubar+g\rightarrow \tbar$ process and top quarks or top antiquarks produced in the $c+g\rightarrow t$ and $\cbar+g\rightarrow \tbar$ processes. The two discriminants $D_1$ and $D_2$ exploit these differences.
 
The first network was trained only with events from the $cgt$ process and was thus optimised for events featuring a sea quark or antiquark in the initial state. The discriminant obtained from this network is defined to be $D_1$. The second NN was trained with events from top-quark production via the $ugt$ process as signal, excluding the charge-conjugate process of top-antiquark production. The corresponding discriminant is called $D_2$. The discriminant $D_1$ is used in a search for the $cgt$ process. The second analysis searches for the $ugt$ process and makes use of both discriminants, $D_1$ and $D_2$. In this case, two SRs were defined based on $\sgn q(\ell)$. The discriminant $D_1$ was used in the $\ell^-$ channel targeting top-antiquark production ($\ubar+g\rightarrow \tbar$). The discriminant $D_2$ was computed in the $\ell^+$ channel, aimed at the search for direct top-quark production ($u+g\rightarrow t$).
 
The NNs were implemented using the NeuroBayes package~\cite{feindt-2004,Feindt:2006pm}, which combines a three-layer feed-forward NN with a complex and robust preprocessing of the input variables before they are presented to the NN. The training of the NNs was based on generated signal and background events and used back-propagation to determine the weights of connections among nodes. As a non-linear activation function, NeuroBayes uses the symmetric sigmoid function
\[ S(x) = \frac{2}{1+e^{-x}}-1 \]
which maps the interval $(-\infty, +\infty\,)$ to the interval $(-1,+1\,)$. In the region close to zero, the sigmoid function has a linear response. The $D_1$ and $D_2$ discriminants were obtained by linearly scaling the outputs of the corresponding NNs to the interval $(0,1)$.
 
Sets of input variables were selected based on studies considering the sensitivity of the analyses as given by the expected upper limits on the production cross-sections (Section~\ref{sec:result} provides more details about the computation of upper limits), how well the observed distributions of the input variables are modelled by simulation, and the ranking of the input variables provided by the preprocessing step of NeuroBayes. The $D_1$ NN used 12 input variables, the $D_2$ NN nine. Six of those variables were common to both NNs. Table~\ref{tab:InputVariables} provides the list of input variables.
\begin{table}[!tbh]
\begin{center}
\caption{Input variables to the two NNs.\label{tab:InputVariables}}
\begin{tabular}{ll}
\toprule
Variable & Definition\\ \midrule
\multicolumn{2}{c}{Variables common to the $D_1$ and $D_2$ NNs} \\ \midrule
$\pT(b)$     & Transverse momentum of the $b$-tagged jet. \\
$m(\ell b)$  & Invariant mass of the charged lepton ($\ell$) and the $b$-tagged jet ($b$). \\
$\mtw$           & Transverse mass of the reconstructed $W$ boson. \\
$\Delta R (W, b)$ &  Distance in the $\eta$--$\phi$ plane between the reconstructed $W$ boson and the $b$-tagged jet. \\
$|\Delta\phi(W, b)|$ & Azimuthal angle between the reconstructed $W$ boson and the $b$-tagged jet. \\
$m(\ell \nu b)$  & Top-quark mass reconstructed from the charged lepton, neutrino, and $b$-tagged jet. \\ \midrule
\multicolumn{2}{c}{Variables used only for the $D_1$ NN} \\ \midrule
$\sgn q(\ell)$  & Sign of the charge of the primary lepton. \\
$\HT(\ell, b, \MET)$ & Scalar sum of the transverse momenta of all reconstructed objects. \\
$\eta(W)$   & Pseudorapidity of the reconstructed $W$ boson. \\
$|\Delta\phi(\ell,\vecpmiss)|$ & Azimuthal angle between the charged lepton and \vecpmiss. \\
$|\Delta\phi(W, \ell)|$ & Azimuthal angle between the reconstructed $W$ boson and the charged lepton. \\
$\pT(\ell \nu b)$  & Transverse momentum of the reconstructed top quark. \\ \midrule
\multicolumn{2}{c}{Variables used only for the $D_2$ NN} \\ \midrule
$\eta(b)$ & Pseudorapidity of the $b$-tagged jet. \\
$\pT(W)$  & Transverse momentum of the reconstructed $W$ boson. \\
$\Delta R (\ell\nu b, W)$ & Distance in the $\eta$--$\phi$ plane between the reconstructed top quark and $W$ boson. \\ \bottomrule
\end{tabular}
\end{center}
\end{table}
Some of the variables, for example $\Delta R (W, b)$ and $m(\ell \nu b)$, required reconstruction of the leptonically decaying $W$ boson, which in turn required reconstruction of the neutrino momentum. While the $x$- and $y$-components of the neutrino momentum, $p_x(\nu)$ and $p_y(\nu)$, were approximated by the components of \vecpmiss, the $z$-component, $p_z(\nu)$, was determined by constraining the mass of the reconstructed $W$ boson to match the measured world average. If the resulting quadratic equation had two real solutions, the one with the smallest $|p_z(\nu)|$ was chosen. In the case of complex solutions, which could occur due to the limited \MET resolution, a kinematic fit was performed that rescaled the neutrino $p_x$ and $p_y$ such that the imaginary part vanished and at the same time the transverse components of the neutrino momentum were kept as close as possible to the \vecpmiss. The $W$ boson was formed by adding the four-vectors of the reconstructed neutrino and the charged lepton.
 
NeuroBayes uses Bayesian regularisation techniques for the training process to improve the generalisation performance and to avoid overtraining. In general, the network infrastructure consists of one input node for each input variable plus one bias node, an arbitrary, user-defined number of hidden nodes arranged in a single hidden layer, and one output node which gives a continuous output in the interval $(-1,+1)$. For the two NNs of this analysis, 15 nodes were used in the hidden layer and the ratio of signal to background events in the training was chosen to be 1:1. The different background processes were weighted according to their expected number of events. Only \ttbar, $W$+jets and single-top-quark events were used as background processes in the training. The multijet background was not used, since its modelling has considerable uncertainties and attempting to optimise the separation of this background from signal events would likely make the results of the analysis more sensitive to any mismodelling of the kinematics of multijet production. After the training step, samples of simulated signal and background events as well as the observed events were processed by the NNs. The resulting distributions of $D_1$ and $D_2$ normalised to unit area are shown in Figure~\ref{fig:nnshape_SR}.
\begin{figure}[!t]
\begin{center}
\subfloat[]{
\includegraphics[width=0.45\textwidth]{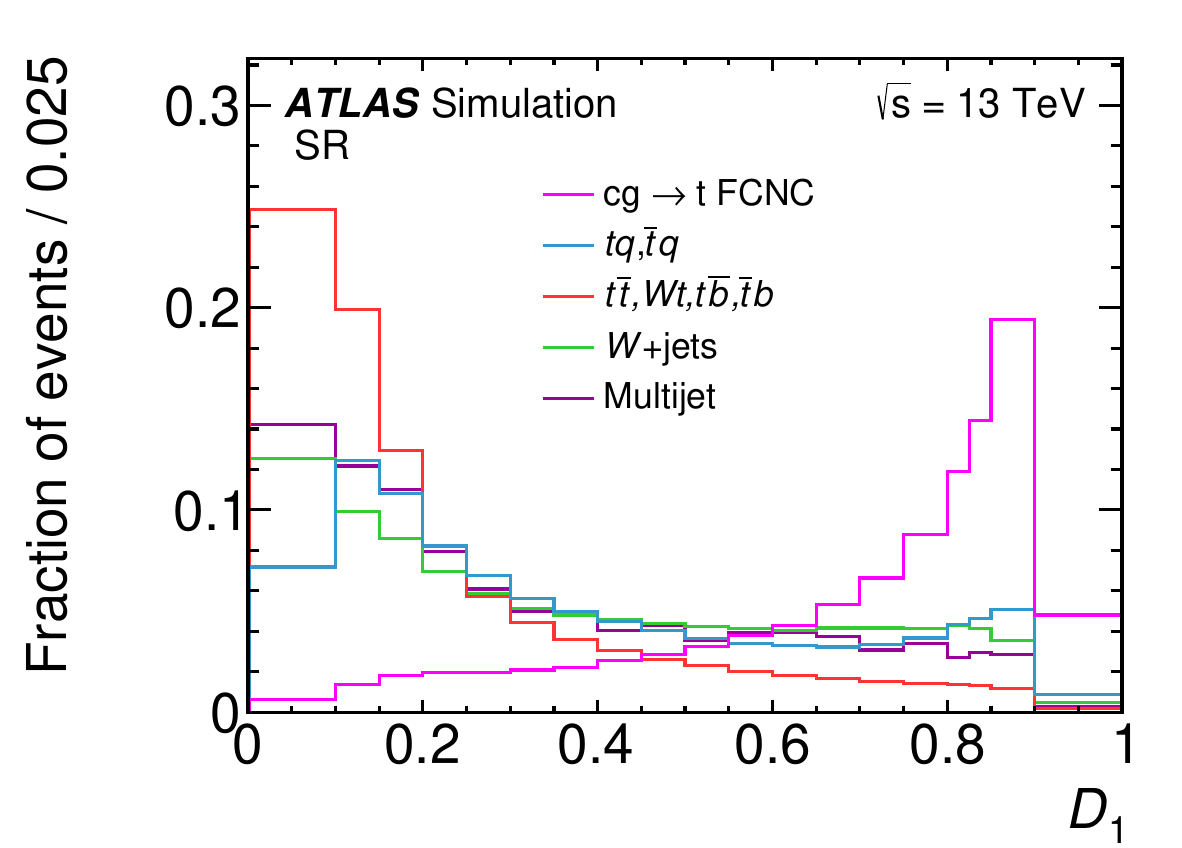}
\label{subfig:D1_cgt}}\qquad
\subfloat[]{
\includegraphics[width=0.45\textwidth]{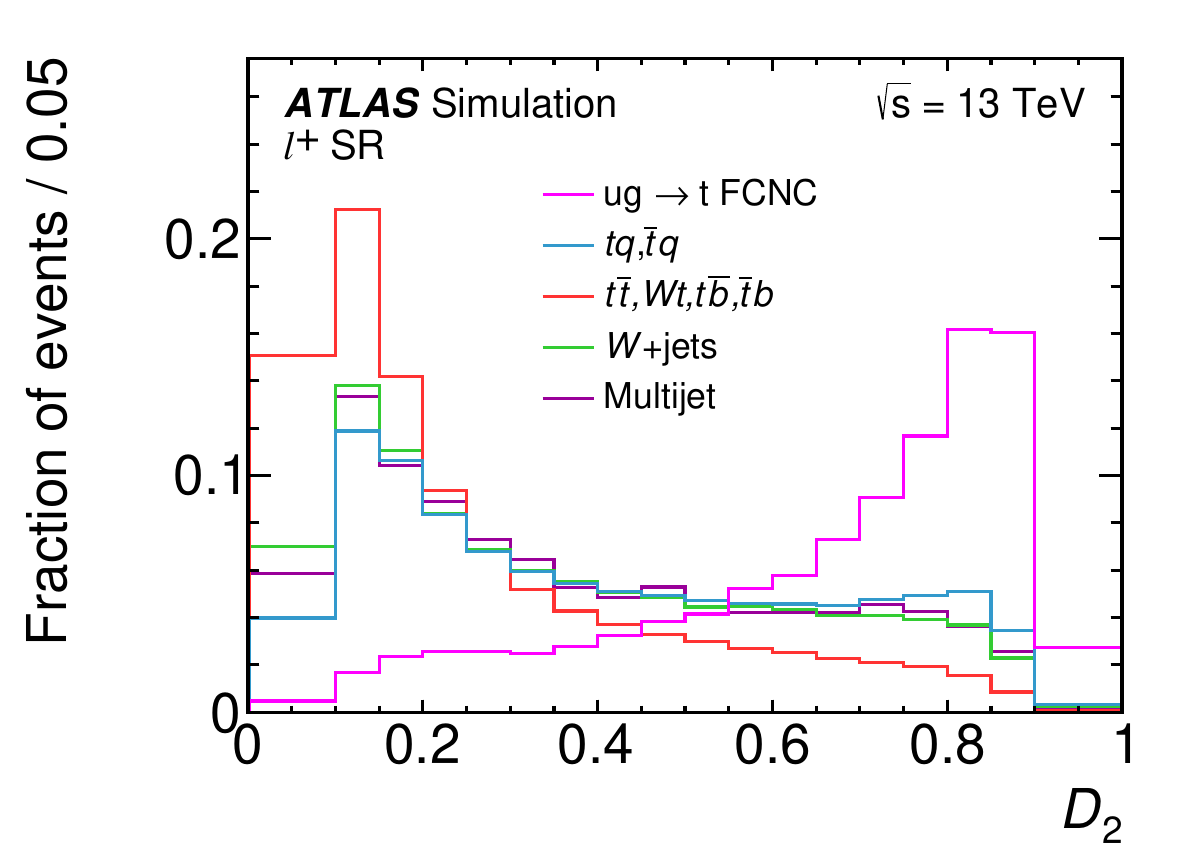}
\label{subfig:D2_ugt_plus}}
\caption[]{Distributions of the $D_1$ and $D_2$ discriminants for signal and background processes. Each distribution is normalised to unit area. The discriminant $D_1$ was used for $cgt$ analysis and the $\ell^-$ channel of the $ugt$ analysis.
The discriminant $D_2$ was used in the $\ell^+$ channel of the $ugt$ analysis. The histograms in~\subref{subfig:D1_cgt} show the distributions obtained in the $cgt$ analysis, that is, the discriminant was evaluated for all selected events independent of $\sgn q(\ell)$. The distributions in the $\ell^+$ channel of the $ugt$ analysis are shown in~\subref{subfig:D2_ugt_plus}.}
\label{fig:nnshape_SR}
\end{center}
\end{figure}
The signal distributions peak at high values between $0.8$ and $0.9$, while the distributions of the background processes peak at low values. Compared to the $t\bar{t}$ process, which has a low event fraction in the highest bins, the $tq$ and $W$+jets production processes have higher event fractions in the most signal-like bins.

Prior to the application of the NNs to the observed collision data in the SR, the modelling of the input variables was checked. The corresponding distributions in the VRs were validated as well. The normalisation of the different scattering processes in the grouping reported in \Fig{\ref{fig:pie_chart_SR}} was taken from the fits to the $\met$ and $\mtw$ distributions for the estimation of the multijet background, reported in Section~\ref{sec:multijet}. As an additional check, the trained NNs were applied in the VRs using input variables corresponding to those in the SR. Three examples of discriminant distributions in the VRs are presented in Figure~\ref{fig:nnCRStack}. In all cases, the model describes the observed discriminant distributions within the estimated uncertainties.
 
\begin{figure}[tbh]
\begin{center}
\subfloat[]{
\includegraphics[width=0.45\textwidth]{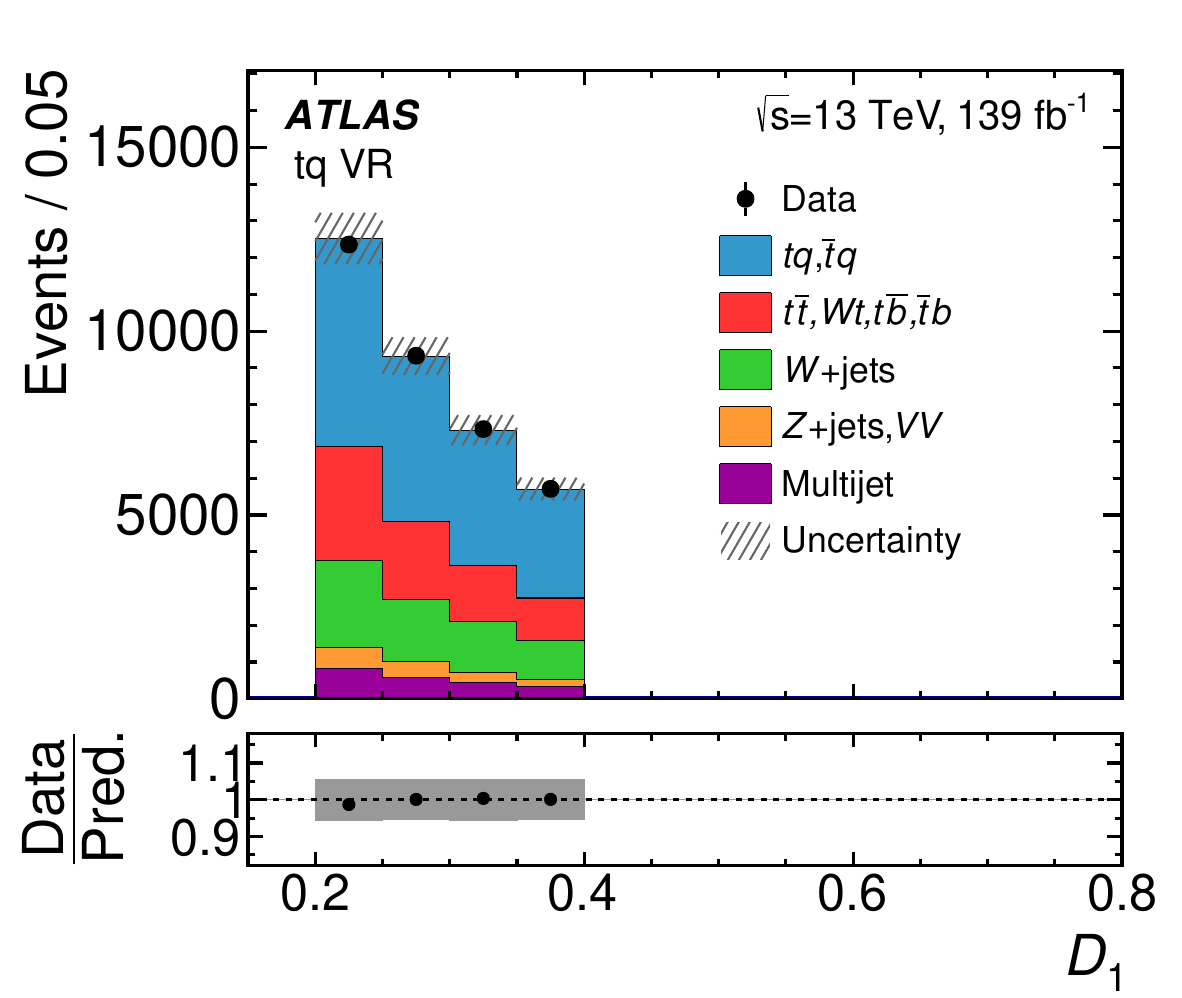}
\label{subfig:D_1_tqVR}
}\qquad
\subfloat[]{
\includegraphics[width=0.45\textwidth]{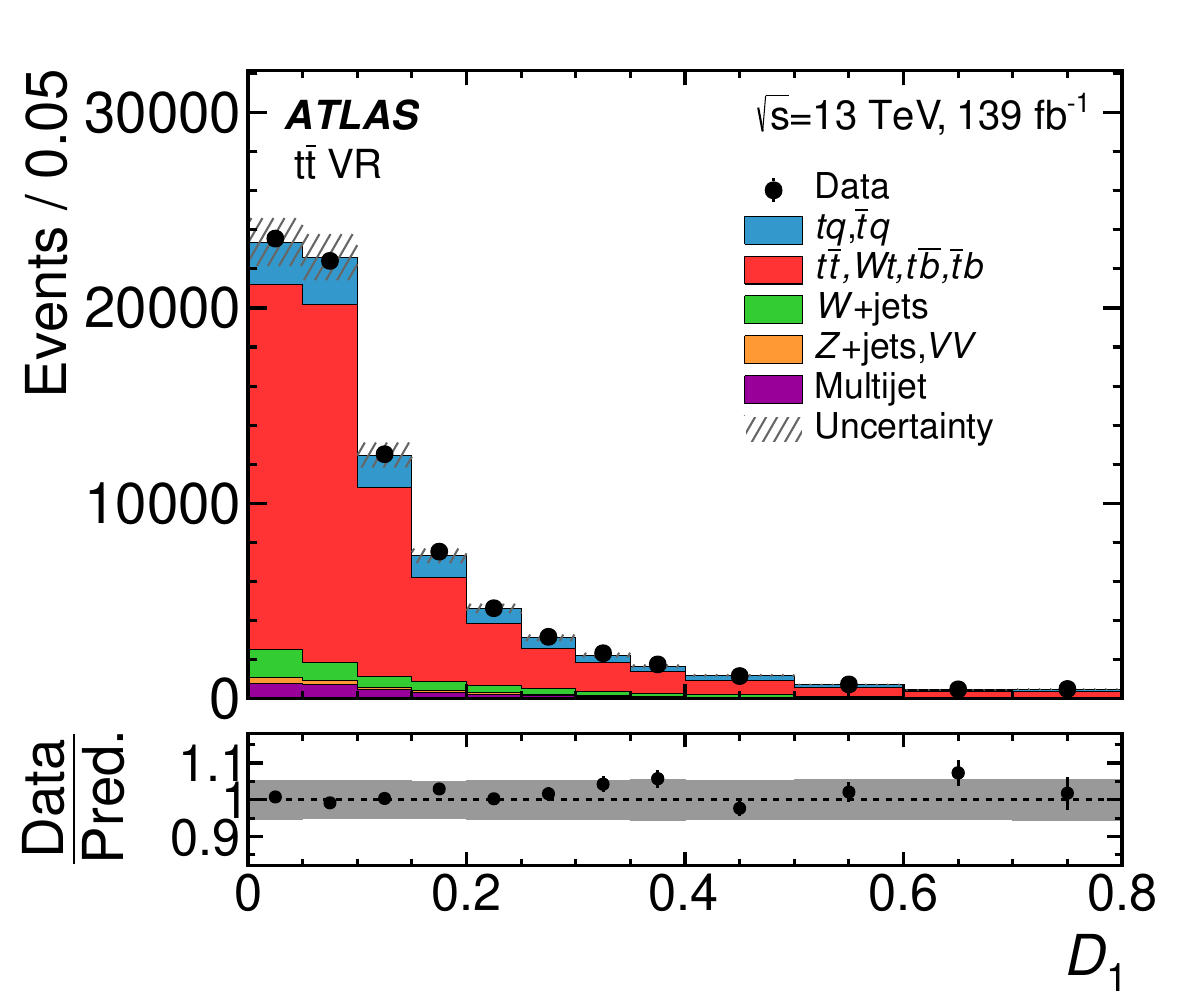}
\label{subfig:D_1_tTbarVR}
}
\newline
\subfloat[]{
\includegraphics[width=0.45\textwidth]{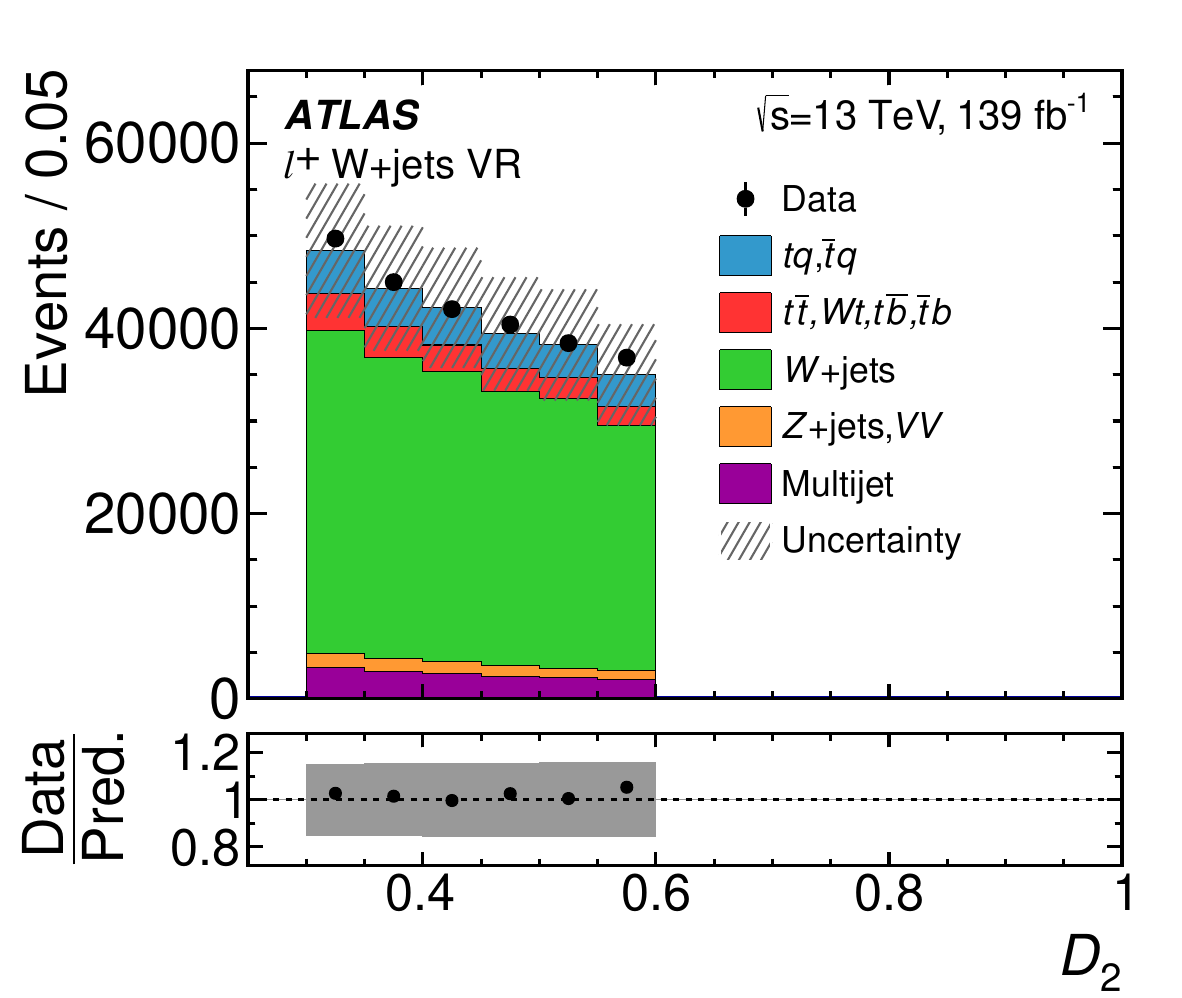}
\label{subfig:D_2_WjetsVR}
}
\caption[]{\label{fig:nnCRStack}Distributions of the NN discriminants $D_1$ and $D_2$ observed in~\subref{subfig:D_1_tqVR}
the \tchanvr, \subref{subfig:D_1_tTbarVR} the \ttvr and \subref{subfig:D_2_WjetsVR} the $\ell^+\;$\wvr. The observed data are compared with the distribution
of simulated background events normalised to the number of expected events, accounting for the normalisation factors obtained
in the fits of the $\met$ and $\mtw$ distributions for estimating the multijet background. The uncertainty band represents the
uncertainty due to limited sample size and the rate uncertainties of the different processes ($20$\% for $W$+jets production, $30$\% for the multijet background and $6$\% for the top-quark processes).
The ratio of observed to predicted (Pred.) numbers of events in each bin is shown in the lower panel.}
\end{center}
\end{figure}
 
\section{Systematic uncertainties}
\label{sec:systematics}
Several sources of systematic uncertainty affect the expected event yield from signal and background processes as well as the shape of the NN discriminants used in the maximum-likelihood fit. The systematic uncertainties are divided into two major categories. Experimental uncertainties are associated with the reconstruction of the four-momenta of final-state partonic objects: electrons, muons, $b$-jets, and $\MET$ as an indication of a primary neutrino. The second category of uncertainties is related to the modelling of scattering processes with event generators.
In the following, the estimation of experimental and modelling uncertainties is explained in more detail.
 
\subsection{Experimental uncertainties}
The uncertainty in the integrated luminosity of the combined 2015--2018 data set is 1.7\% and is based on a calibration of the luminosity scale using $x$--$y$ beam-separation scans~\cite{ATLAS-CONF-2019-021}. The luminosity uncertainty was applied to the signal and background event yields except for the multijet background, which was estimated in a data-driven way.
Scale factors were applied to simulated events to correct for reconstruction, identification, isolation and trigger performance differences between data and detector simulation for electrons and muons.
These scale factors, as well as the lepton momentum scale and resolution, were assessed using $Z\rightarrow \ell^+\ell^-$ events in simulation and data~\cite{EGAM-2018-01,Aad:2020gmm}. Their systematic uncertainties were propagated to the expected event yields and discriminant distributions used in the maximum-likelihood~fit.
 
The jet energy scale (JES) was calibrated using a combination of test-beam data, simulation and in situ techniques~\cite{PERF-2016-04}. Its uncertainty is decomposed into a set of 30 uncorrelated components, of which 29 are non-zero in a given event depending on the type of simulation used. Sources of uncertainty contributing to the JES uncertainties include pile-up modelling, jet flavour composition, single-particle response and effects of jets not fully contained within the calorimeter.
The uncertainty of the jet energy resolution (JER) is represented by eight components accounting for jet-$\pt$ and $\eta$-dependent differences between simulation and data~\cite{PERF-2011-03}. The uncertainty in the efficiency to pass the JVT requirement for pile-up suppression was also considered~\cite{PERF-2014-03}.
 
The uncertainties in the $b$-tagging calibration were determined for $b$-jets~\cite{FTAG-2018-01}, broken down into 45 orthogonal components. The uncertainties depend on the \pt of the $b$-jets and were propagated through the analysis as weights. Since $b$-jets were identified with very high purity, the misidentification rate of $c$-jets and light-flavour jets was very low and a dedicated calibration was not performed.
Only the $W+$jets background has a small component of misidentified $c$-jets and light-flavour jets. For other backgrounds and for the signal processes these components are negligible. Since the rate of the $W+$jets background was determined directly from the final maximum-likelihood fit, there was no need for an overall rate uncertainty on the $W+$jets background. Instead a dedicated shape uncertainty was assigned to the modelling of the contamination by $c$-jets and light-flavour jets. More details are given in the next section on modelling uncertainties.
 
The uncertainty in \met due to a possible miscalibration of its soft-track component was derived from data--simulation comparisons of the \pT balance between the hard and soft \met components~\cite{PERF-2016-07}. To account for pile-up distribution differences between simulation and data, the pile-up profile in the simulation was corrected to match the one in data. The uncertainty associated with the correction factor was applied.

\subsection{Modelling uncertainties}
Uncertainties in the theoretical cross-sections were evaluated for the SM top-quark processes ($tq$, $t\tbar$, $tW$ and $t\bbar$) as quoted in Section~\ref{sec:samples_top_quark}.
The single largest background, $W+$jets production, was allowed to float in the likelihood fit and thus a cross-section uncertainty was not applied.
A symmetric uncertainty of $\pm 20$\% was assigned to the $Z+$jets production cross-section by evaluating the effect of seven variations of $\muR$ and $\muF$ in the matrix-element computation~\cite{Bothmann:2016nao}. In this estimate, which is meant to account for missing higher-order corrections, the scales were independently varied by factors of $0.5$ and $2.0$, avoiding the variations with ratios of four between the two scales. The biggest impact on the cross-sections was found for a correlated variation of $\muR$ and $\muF$. The same uncertainty of $\pm 20$\% was assigned to diboson production. The uncertainty in the event yield of the multijet background is $30$\%.
 
Uncertainties in modelling parton showers and hadronisation were assigned to the FCNC signal and the SM top-quark production processes (\ttbar, $tW$, $tq$ and $t\bar{b}$ production) by comparing the nominal samples with alternative samples for which \METOP and \POWHEGBOX were interfaced to \HERWIG[7.0.4] instead of \PYTHIA[8.235] or \PYTHIA[8.230], respectively. When generating parton showers the \textsc{MMHT2014lo}~\cite{Harland-Lang:2014zoa} PDF set was used as well as the H7-UE-MMHT~\cite{Bellm:2015jjp} set of tuned parameters. The uncertainties were defined independently for each scattering process, namely the FCNC signal process and the four SM top-quark production processes. In addition, normalisation and shape effects were decorrelated as well.
 
Uncertainties related to the choice of renormalisation and factorisation scales for the matrix-element calculations were evaluated by varying the scales in a correlated way by factors of $2$ and $0.5$, separately for each process. In the case of the FCNC signal processes, dedicated samples of simulated events were generated with varied scales. For the SM top-quark production processes and for \wjets production, the scale variations were implemented as generator weights in the nominal sample. These weights were propagated through the entire analysis.
 
The uncertainty due to the choice of a scale for matching the matrix-element calculation of the $t\tbar$ process to the parton shower was estimated using an additional \ttbar sample produced with the $h_\mathrm{damp}$ parameter set to $3\times m_t$, while keeping all other generator settings the same as for the nominal sample of $\ttbar$ events. The uncertainty due to the choice of matrix-element-to-parton-shower matching scale used in the generation of the FCNC signal samples was evaluated by comparisons with alternative samples produced with a matching scale of $15\,\GeV$ instead of the $10\,\GeV$ scale used for the nominal sample.
The uncertainty related to the specific algorithm for matching the NLO-matrix-element computation to parton showers was evaluated for the SM top-quark production processes ($tq$, $t\tbar$, $tW$ and $t\bbar$) by comparing samples generated by \POWHEGBOX with samples generated by \MGNLO~\cite{Alwall:2011uj}. Both set-ups used \PYTHIA for the parton-shower computation. The effects of this matching-algorithm uncertainty on the shape of the NN discriminants and on the event yields were decorrelated in the maximum-likelihood fit.
 
Uncertainties in the amount of initial-state and final-state radiation were assessed for the FCNC signal processes and the SM top-quark production processes by varying the parameter \texttt{Var3c} of the A14 parton-shower tune within the uncertainties of the tune and, for final-state radiation, by varying the renormalisation scale $\muR$, at which the strong coupling constant $\alpha_\mathrm{s}$ was evaluated, by factors of $0.5$ and $2.0$. The two variations, the one of \texttt{Var3c} and the one of $\muR$, were handled independently.
The uncertainty due to the scheme for removing the overlap of the $tW$ process with $\ttbar$ production was evaluated by comparing the nominal sample, using the diagram-removal scheme, with a sample produced with an alternative scheme (diagram subtraction)~\cite{Frixione:2008yi}.
In all uncertainty evaluations mentioned above the alternative samples or reweighted samples were normalised to the total cross-section of the nominal samples.
 
Uncertainties due to PDFs were evaluated for the $tq$ process and the combined $t\tbar$-$tW$-$t\bbar$ process using the PDF4LHC15 combined PDF set~\cite{Butterworth:2015oua} with 30 symmetric eigenvectors. Samples of simulated events were reweighted to the central value and the eigenvectors of the combined PDF set. Systematically varied templates were constructed by taking the differences between the samples reweighted to the central value and those reweighted to the eigenvectors. In the likelihood fit the PDF uncertainties were treated as correlated between the $tq$ process and the combined $t\tbar$-$tW$-$t\bbar$ process.
The uncertainty in the average number of interactions per bunch crossing was accounted for by varying accordingly the scale factors applied to weight the simulated events in order to obtain the pile-up distribution observed in collision data.
 
The uncertainty in the multijet background was evaluated by modifying the respective selection criteria for the jet-lepton and the non-prompt-muon candidate. For each lepton type, two alternative selections were defined by varying the requirements on the energy fraction measured in the electromagnetic calorimeter in the case of the jet-lepton and by varying the isolation criteria for the muon candidates. The variations leading to the larger deviations from the nominal set-up were chosen when defining uncertainties in the shape of the NN discriminant distribution for the multijet background.
 
With a fraction of $92$\% the $W$+$b$-jets component dominates the $W+$jets background. Since the number of simulated events with jets of different flavour, $c$-jets or light-flavour jets, was very limited, the $W+$jets template was based on the $W$+$b$-jets component only. The expected event yield was scaled such that the events with jets of different flavour were also considered.  To account for small shape differences between the NN-discriminant distributions for $W$+$b$-jets, $W$+$c$-jets and $W+$light-flavour jets, two alternative $W+$jets template histograms were created by adding to the nominal $W$+$b$-jets component the $W$+$c$-jets and $W+$light-flavour jets contributions with three times the expected rate.
The resulting shape differences were applied in a symmetric way in the maximum-likelihood fit, which constrained the input uncertainties to a level of $80$\% for $W$+$c$-jets in both searches ($ugt$ and $cgt$) and $40$\% ($70$\%) for $W+$light-flavour jets in the $ugt$ analysis ($cgt$ analysis).
In the $cgt$ analysis using the $D_1$ discriminant, the $W$+$c$-jets input uncertainty is approximately $\pm 9$\% for NN discriminant values below $0.7$ and rises to $\pm 12$\% at high values. The $W+$light-flavour jets input uncertainty is approximately $\pm 4$\% for NN discriminant values of $0.75$ to $1.0$, while at values below $0.1$ it reaches $\pm 10$\%. In the $ugt$ analysis, the corresponding uncertainties in the shapes of the $D_1$ and $D_2$ discriminants have very similar features.
 
The uncertainties due to the finite number of simulated events, also called the MC statistical uncertainty, was accounted for by adding a nuisance parameter for each bin of the NN discriminant distributions separately for each scattering process, implementing the Barlow--Beeston approach~\cite{Barlow:1993dm}.
 
\section{Results}
\label{sec:result}
The observed distributions of the NN discriminants were subjected to a binned maximum-likelihood fit, probing for a potential FCNC signal. Two analyses were performed, searching separately for the $ugt$ and $cgt$ FCNC processes.
The likelihood function $\mathcal{L}$ was constructed as a product of Poisson probability terms over all bins of the NN discriminants. The function $\mathcal{L}(\mu, \vec{\theta})$ depends on the signal-strength parameter $\mu$, a multiplicative factor to the signal production cross-section used to normalise the simulated signal samples, and $\vec{\theta}$, a set of nuisance parameters including the effects of systematic uncertainties on the signal and background expectations. The range of each nuisance parameter was constrained in the likelihood function by a Gaussian term. The signal strength of the $W+$jets process, $\mu(Wj)$, was treated as a free multiplicative factor as well. In the $ugt$ analysis, in contrast to the $cgt$ analysis, the rates of the $W^++$jets and $W^-+$jets processes were determined separately in a simultaneous fit using two independent normalisation parameters.
 
Systematically varied discriminant distributions were smoothed and nuisance parameters of systematic uncertainties with negligible impact were entirely removed in order to reduce spurious effects in minimisation, improve convergence of the fit, and reduce the computing time. Normalisation and shape effects of a source of systematic uncertainty were treated separately in the pruning process.
 
Single-sided systematic variations were turned into symmetric variations by taking the full difference in event yield and shape between the nominal model and the alternative model and mirroring this difference in the opposite direction. For sources with two variations, their effects were made symmetric by using the average deviation from the nominal prediction.
 
\subsection{Results of the profile likelihood fit}
The results of the maximum-likelihood fits yielded: $\mu(ugt)=0.10\pm 0.18$\footnote{The reference cross-sections of the signal processes are $\sigma(u+g\rightarrow t)\times\mathcal{B}(t\rightarrow Wb)\times\mathcal{B}(W\rightarrow \ell\nu)=6.27\,$pb and $\sigma(c+g\rightarrow t)\times\mathcal{B}(t\rightarrow Wb)\times\mathcal{B}(W\rightarrow \ell\nu)=8.52\,$pb.}, $\mu(W^+j)=1.25\pm 0.15$ and $\mu(W^-j)=1.32\pm 0.17$
as well as $\mu(cgt)=0.15\pm 0.17$ and $\mu(Wj)=1.19\pm 0.15$.
The normalisation factors of the $W+$jets process were determined to be above $1$ in all cases. This finding is similar to that from a dedicated measurement of the cross-section of the $Z+{\geq}1$ $b$-jet process~\cite{STDM-2017-38}. No significant nuisance parameters pulls were observed in either fit. The impact of systematic uncertainties on the sensitivity is much larger than that of the data statistical uncertainties.
An importance ranking of systematic uncertainties was determined by computing the shifts $\Delta\mu_j$ in the signal-strength parameters $\mu(ugt)$ and $\mu(cgt)$ when fixing a particular nuisance $\theta_j$ related to the uncertainty $j$ to its estimated value $\hat{\theta}_j$ plus or minus its post-fit uncertainty $\Delta\hat{\theta}_j$. In the ranking, the uncertainties are sorted in $|\Delta\mu_j|$ in descending order.
 
The five leading systematic uncertainties in the $ugt$ fit are due to the MC statistical uncertainty in the highest bin of the NN discriminant $D_2$ of the $W+$jets process, the $W$+$c$-jets shape uncertainty, the first effective nuisance parameter of the uncertainty in the jet-energy resolution, the MC statistical uncertainty in the second-highest bin of the NN discriminant $D_2$ of the $W+$jets process, and the normalisation component of the uncertainty in the matrix-element-matching algorithm of the $tq$ process.
Out of these leading uncertainties, the three non-MC-statistical uncertainties were constrained in the fit to the range of 80\% to 90\% of their original value.
The five leading systematic uncertainties in the $cgt$ fit are due to the modelling of the parton shower of the FCNC $cgt$ process, the shape component of the parton-shower uncertainty of the $tq$ process, the uncertainty in the resolution of the soft-track term of the $\MET$ computation, the shape component of the uncertainty in the matrix-element-matching algorithm of the $tq$ process, and the MC statistical uncertainty in the highest bin of the NN discriminant $D_1$ of the $W+$jets process.
Out of these leading uncertainties of the $cgt$ analysis, the fit constrained the three non-MC-statistical uncertainties to the range of 65\% to 90\% of their original value.
 
Table~\ref{tab:res_evnt_yield_final} provides the expected, the observed, and the fitted event yields in the SR.
\begin{table}[!tb]
\begin{center}
\caption{\label{tab:res_evnt_yield_final}Expected pre-fit and post-fit event yields along with the observed event yield in the SR. The quoted uncertainties include the statistical and systematic uncertainties of the event yields. Correlations, including anticorrelations, among the nuisance parameters related to the uncertainties were taken into account as determined in the maximum-likelihood fit.}
\vspace*{3mm}
\begin{tabular}{lSSS}
\toprule
Process & {Pre-fit} & {Post-fit $cgt$} & {Post-fit $ugt$} \\
\midrule
$ugt$ FCNC process       & 0                &  0               & 1200 \pm 2100 \\
$cgt$ FCNC process       & 0                &   4100 \pm 4500  & 0 \\
$tq$                     & 138600 \pm 9300  & 149200 \pm 9400  &  150000 \pm  10000 \\
$t\tbar$, $tW$, $t\bbar$ & 179000 \pm 17000 & 179000 \pm 14000 &  175200 \pm   9700 \\
$W+$jets                 & 229000 \pm 30000 & 281000 \pm 21000 &  292000 \pm  18000 \\
$Z+$jets, $VV$           &  29700 \pm  6000 &  30000 \pm  6000 &   29800 \pm   6000 \\
Multijet                 &  47000 \pm 14000 &  45000 \pm 14000 &   40000 \pm  12000 \\
\midrule
Total                    & 650000 \pm 46000 & 688600 \pm 2400 &  688700  \pm 3500 \\
\midrule
Observed                 & 688380 & 688380 & 688380 \\
\bottomrule
\end{tabular}
\end{center}
\end{table}
The results of the $ugt$ and $cgt$ analyses differ slightly, but agree well within uncertainties. The event yields after the fit account for pulls of the nuisance parameters.
The fitted discriminant distributions are shown in Figures~\ref{fig:NNout_postfit_ugt} and~\ref{fig:NNout_postfit_cgt} for the $ugt$ and $cgt$ analyses, respectively.
\begin{figure}[!pt]
\centering
\subfloat[]{
\includegraphics[width=0.45\textwidth]{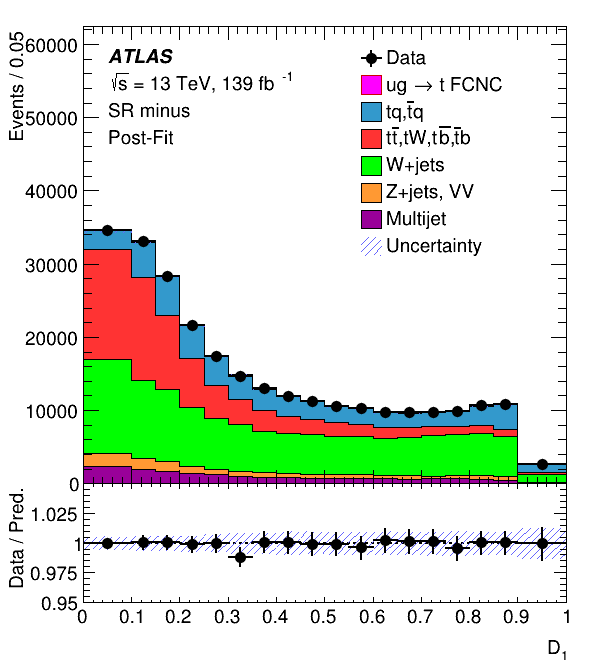}
\label{fig:NNout_postfit_D1_ugt_minus_full_dist}}
\qquad
\subfloat[]{
\includegraphics[width=0.45\textwidth]{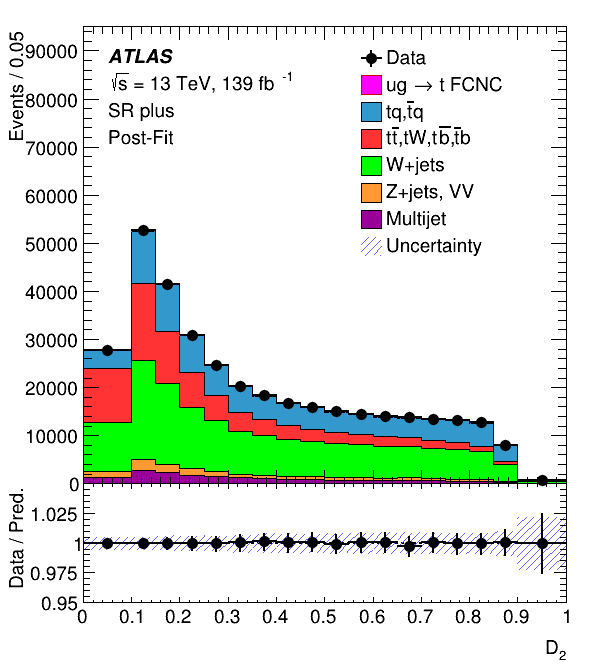}
\label{fig:NNout_postfit_D2_ugt_plus_full_dist}} \\
\subfloat[]{
\includegraphics[width=0.45\textwidth]{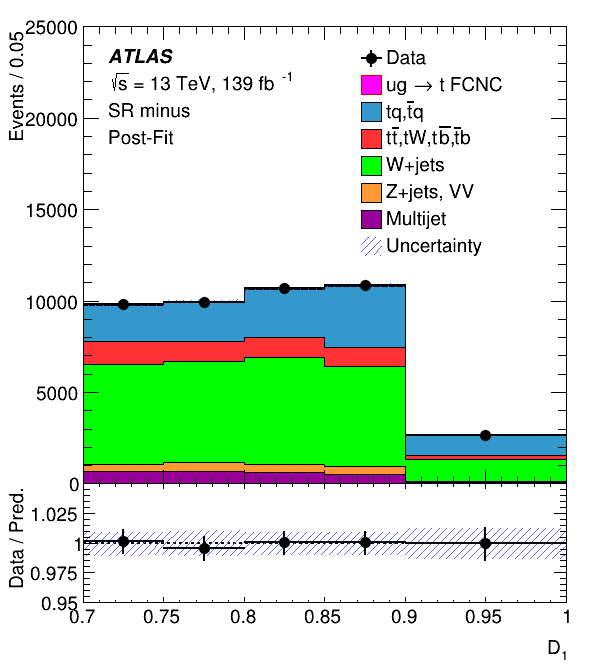}
\label{fig:NNout_postfit_D1_ugt_minus_zoom_high}}
\qquad
\subfloat[]{
\includegraphics[width=0.45\textwidth]{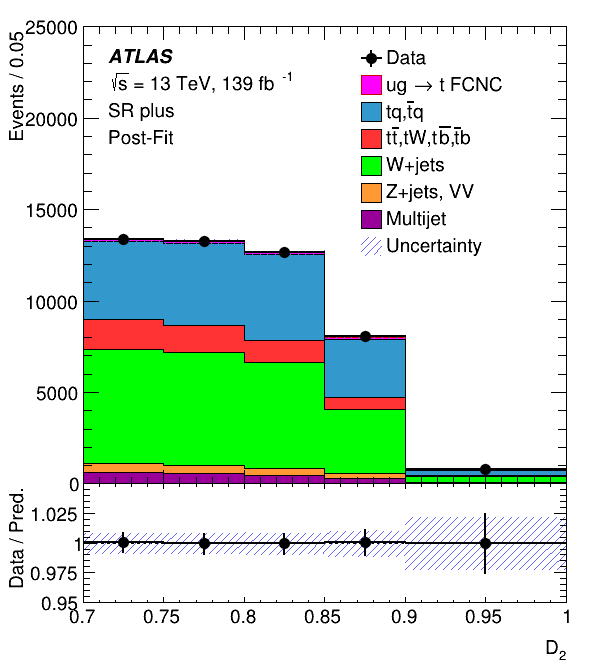}
\label{fig:NNout_postfit_D2_ugt_plus_zoom_high}}
\caption{\label{fig:NNout_postfit_ugt}The NN discriminants $D_1$ and $D_2$ of the $ugt$ search are shown with the post-fit normalisation applied to the stacked histograms of the different hard-scattering processes. The histograms in \protect\subref{fig:NNout_postfit_D1_ugt_minus_full_dist} and \protect\subref{fig:NNout_postfit_D2_ugt_plus_full_dist} show the full discriminant range in the negatively charged lepton channel and the positively charged lepton channel, respectively. The histograms \protect\subref{fig:NNout_postfit_D1_ugt_minus_zoom_high} and \protect\subref{fig:NNout_postfit_D2_ugt_plus_zoom_high} show a zoomed-in view of the high discriminant region between 0.7 and 1.0. The hatched bands represent the post-fit uncertainty of the total event yield in each bin. Correlations among uncertainties were taken into account as determined in the fit. The fitted signal contribution is included but is barely visible because its relative size is very small.}
\end{figure}
\begin{figure}[!h!tpb]
\centering
\subfloat[]{
\includegraphics[width=0.45\textwidth]{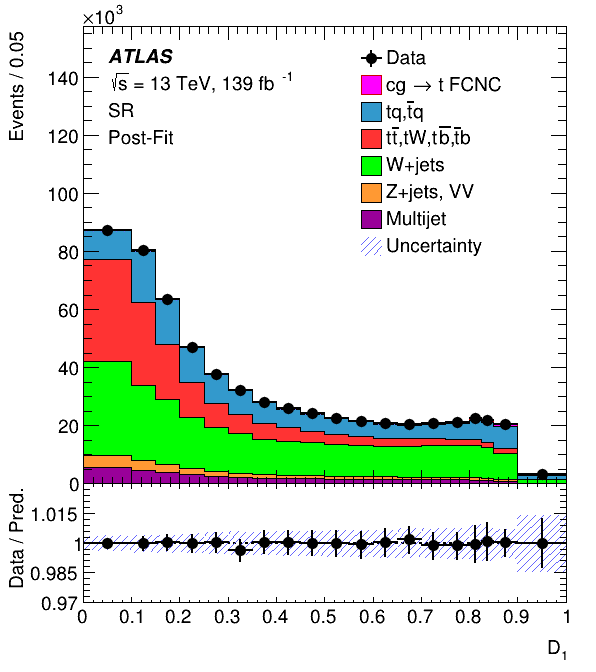}
\label{fig:NNout_postfit_cgt_full_dist}}
\qquad
\subfloat[]{
\includegraphics[width=0.45\textwidth]{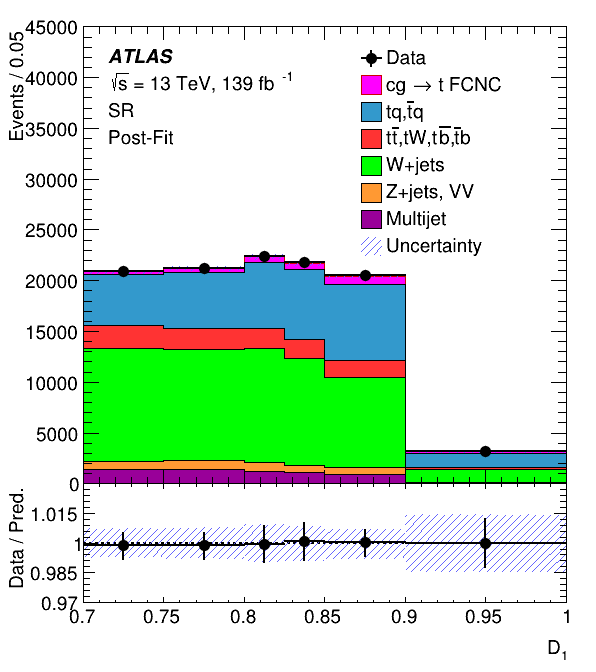}
\label{fig:NNout:postfit_cgt_zoom_high}}
\caption{\label{fig:NNout_postfit_cgt}The NN discriminant $D_1$ of the $cgt$ search is shown with the post-fit normalisation applied to the stacked histograms of the different hard-scattering processes. The histogram in \protect\subref{fig:NNout_postfit_cgt_full_dist} shows the full discriminant range. The histogram \protect\subref{fig:NNout:postfit_cgt_zoom_high} shows a zoomed-in view of the high discriminant region between $0.7$ and $1.0$. The hatched bands represent the post-fit uncertainty of the total event yield in each bin. Correlations among uncertainties were taken into account as determined in the fit. }
\end{figure}
The observed discriminant distributions are very well described by the fitted model and they are compatible with the background-only hypothesis.
 
\FloatBarrier
\subsection{Upper limits on cross-sections, EFT coefficients and branching ratios}
Since the observed NN-discriminant distributions were found to be compatible with the background-only hypothesis, upper limits were set on the cross-sections of the $ugt$ and the $cgt$ processes at the 95\% confidence level (CL). The limits were computed by applying the CL$_\textrm{s}$ method~\cite{0954-3899-28-10-313,JUNK1999435} as implemented in the RooFit package~\cite{RooFit} to the test statistic
\begin{equation}
\tilde{q}_{\mu} = \left\{ \begin{array}{ll} -2 \ln \left(  \frac{\mathcal{L}\left(\mu, \hat{\hat{\vec{\theta}}}(\mu)\right)}{\mathcal{L}\left(0, \hat{\hat{\vec{\theta}}}(0)\right)}\right) &
\mathrm{if}\ \hat{\mu} < 0, \\
-2 \ln \left(  \frac{\mathcal{L}\left(\mu, \hat{\hat{\vec{\theta}}}(\mu)\right)}{\mathcal{L}\left(\hat{\mu}, \hat{\vec{\theta}}\right)}\right) & \mathrm{if}\ 0 \leq \hat{\mu} \leq \mu, \\
\ \ 0 & \mathrm{if}\ \hat{\mu}>\mu. \end{array}  \right.
\label{eq:q_mu}
\end{equation}
In Eq.~(\ref{eq:q_mu}), the symbols $\hat{\mu}$ and $\hat{\vec{\theta}}$ represent the values of the parameters maximising the likelihood function and $\hat{\hat{\vec{\theta}}}$ are the values of the nuisance parameters which maximise the likelihood function for a fixed value of $\mu$. The obtained upper limits on the cross-sections times branching ratio are
\begin{eqnarray}
\sigma(ugt)\times\mathcal{B}(t\rightarrow Wb)\times\mathcal{B}(W\rightarrow \ell\nu) &<&3.0~\mathrm{pb} \ \ \ \ \mathrm{and} \label{eq:xs_limit_u} \\
\sigma(cgt)\times\mathcal{B}(t\rightarrow Wb)\times\mathcal{B}(W\rightarrow \ell\nu) &<&4.7~\mathrm{pb},  \label{eq:xs_limit_c}
\end{eqnarray}
with $\mathcal{B}(W\rightarrow \ell\nu)=0.325$ being the sum of branching ratios of all three leptonic decay modes of the $W$ boson.
The expected cross-section-times-branching-ratio limits are $2.4$~pb and $2.5$~pb, respectively. The observed limits are larger than the expected ones because non-zero signal yields are fitted.
 
The cross-section limits are interpreted within the \texttt{TopFCNC} model~\cite{Degrande:2014tta}, which implements an effective operator formalism and is based on the \textsc{FeynRules\;2.0} framework~\cite{ALLOUL20142250} used inside the \MGNLO event generator.
With this set-up the cross-sections of the FCNC processes under consideration were calculated at NLO in QCD, providing a significant improvement on LO calculations, since NLO corrections for this class of processes were found to be between
30\% and 80\%~\cite{Degrande:2014tta}.\footnote{While \MGNLO can be used for a fixed-order calculation at NLO, events for which a matching to a parton-shower program is needed can only be generated at LO in the current implementation.}
In the \texttt{TopFCNC} model, the two operators $\mathcal{O}_{uG}^{\,ut}$ and $\mathcal{O}_{uG}^{\,ct}$ generate the $ugt$ and $cgt$ processes, and the coupling strengths of the corresponding vertices are given by the two coefficients \CutuG and \CctuG divided by the square of the new-physics scale $\Lambda$.
The total cross-sections are found to be related to the EFT coefficients by
\begin{eqnarray}
\sigma(u+g\rightarrow t) & =& 2773\times \left(\frac{\CutuG}{\Lambda^2}\right)^2\mathrm{pb\,\TeV^{4}}\ \ \ \mathrm{and} \label{eq:FCNC:xs_C_u}\\
\sigma(c+g\rightarrow t) & =& 719\times \left(\frac{\CctuG}{\Lambda^2}\right)^2\mathrm{pb\,\TeV^{4}}. \label{eq:FCNC:xs_C_c}
\end{eqnarray}
Using Eqs.~(\ref{eq:FCNC:xs_C_u}) and~(\ref{eq:FCNC:xs_C_c}) the cross-section limits of Eqs.~(\ref{eq:xs_limit_u}) and~(\ref{eq:xs_limit_c}) become limits on the EFT coefficients:
\begin{equation}
\frac{|\CutuG|}{\Lambda^2} < 0.057\,\TeV^{-2} \ \ \ \ \mathrm{and} \ \ \ \ \frac{|\CctuG|}{\Lambda^2} < 0.14\,\TeV^{-2} \ \ \ \ \mathrm{at~the~95\%~CL.}
\label{eq:EFT_limits}
\end{equation}
Since the $u$-quark is a valence quark of the proton, it carries on average a much larger momentum fraction than
the $c$-quark, and thus the cross-section of the $ugt$ process is much larger
than the cross-section of the $cgt$ process, when considering the same
value of the corresponding coefficient ($\CutuG = \CctuG$).
For a certain experimental sensitivity, the sensitivity to \CutuG is therefore higher
than to \CctuG. However, in the two-Higgs-doublet models mentioned in Section~\ref{sec:intro} the predicted FCNC couplings to charm quarks are much higher than to up quarks. For this reason, the limits on \CctuG have phenomenological relevance even though they are weaker than the limits on \CutuG.
The limits presented in Eq.~(\ref{eq:EFT_limits}) tighten constraints set by the CMS Collaboration using dilepton events recorded in Run 2 of the LHC~\cite{CMS:2019zct} by more than a factor of three. The CMS analysis searched for $tW$ production cross-section via FCNC.
 
An alternative and very accessible way of comparing the upper limits on the EFT coefficient with previous results uses the branching ratios of FCNC top-quark decays: $\mathcal{B}(t\rightarrow u+g)$ and $\mathcal{B}(t\rightarrow c+g)$. These branching ratios are given as a function of the EFT coefficients by the relation
\begin{equation}\nonumber
\mathcal{B}(t\rightarrow q+g)=0.0186 \times \left(\frac{\CqtuG}{\Lambda^2}\right)^2\TeV^4,
\label{eq:EFT_coeff_BR_relation}
\end{equation}
with $q=u,c$~\cite{Durieux:2014xla}, assuming
the top-quark width to be $\Gamma_t=1.32~\GeV$. The resulting upper limits at the 95\% CL are
\begin{equation}
\mathcal{B}(t\rightarrow u+g)<\num{0.61E-4} \ \ \ \ \mathrm{and} \ \ \ \ \mathcal{B}(t\rightarrow c+g)<\num{3.7E-4}.
\label{eq:BR_limits_obs}
\end{equation}
These new bounds are approximately a factor of two more restrictive than the previous ATLAS results obtained at a centre-of-mass energy of $8\,\TeV$~\cite{TOPQ-2014-13-witherratum}. The bound on the $cgt$ mode is comparable to that of the CMS analysis combining 7 and 8~\TeV~data~\cite{CMS_FCNC_8TeV}, while the bound on the $ugt$ mode is significantly weaker than the CMS one.
 
\subsection{Comparison of expected upper limits}
For assessing the sensitivity of this analysis and comparing it with the sensitivity of other results, and for evaluating the impact of different groups of systematic uncertainties, the computation of expected upper limits is more suitable than using the observed results, since biases caused by statistical fluctuations are avoided and the signal contribution is set to zero. The expected limits were derived by using the expected distributions of the NN discriminants, considering background processes only. The initially predicted rate of the $W+$jets process was scaled by a factor of $\mu(W^+b)=1.22$ or $\mu(W^-b)=1.30$ for the $ugt$ analysis and by a factor of $\mu(Wb)=1.18$ for the $cgt$ analysis. These normalisation factors were obtained from background-only fits to the observed NN discriminants in background-dominated regions, namely the ranges $0.0$ to $0.7$ for the $D_1$ discriminant and $0.0$ to $0.55$ for the $D_2$ discriminant. The resulting expected upper limits in terms of branching ratios are
\begin{equation}
\mathcal{B}_{95}^\mathrm{exp}(t\rightarrow u+g)=\num{0.49E-4} \ \ \ \ \mathrm{and} \ \ \ \ \mathcal{B}_{95}^\mathrm{exp}(t\rightarrow c+g)=\num{2.0E-4}.
\label{eq:BR_limits_exp}
\end{equation}
Compared to the ATLAS analysis at 8~\TeV\ centre-of-mass energy, significant improvements in sensitivity are obtained for both the $ugt$ and $cgt$ analyses. However, the improvements in sensitivity are smaller than expected from a simple scaling of the number of expected events with the increase in integrated luminosity and the increase in signal cross-sections. The main reason for this effect is that the cross-sections of the top-quark background processes rise faster with the centre-of-mass energy than the cross-sections of the FCNC signal processes.
 
The expected upper limits are lower than the observed upper limits in Eq.~(\ref{eq:BR_limits_obs}), since non-zero, yet insignificant, signals are observed, while the expected limits are obtained from expected distributions without any signal events included. The effect is larger for the $cgt$ analysis than for the $ugt$ analysis because the fitted signal event yield is more than three times larger in the $cgt$ case, as seen in Table~\ref{tab:res_evnt_yield_final}.
 
In order to quantify the impact of different groups of systematic uncertainties, expected upper limits were computed for different scenarios:
(1) include only data statistical uncertainties, (2) include the experimental systematic uncertainties in addition, (3) include all systematic uncertainties except for the MC statistical uncertainties and (4) include all uncertainties. The last case leads to the limits quoted in Eq.~(\ref{eq:BR_limits_exp}). The results of this study are reported in Table~\ref{tab:syst_unc_impact}
\begin{table}[!tbp]
\begin{center}
\caption{\label{tab:syst_unc_impact}Impact of systematic uncertainties on the expected upper limits on the branching ratios of the FCNC decay modes $\mathcal{B}(t\rightarrow u+g)$ and $\mathcal{B}(t\rightarrow c+g)$. Four scenarios are considered: (1) include only data statistical uncertainties, (2) include the experimental systematic uncertainties in addition, (3) include all systematic uncertainties except for the MC statistical uncertainties and (4) include all uncertainties.}
\vspace*{3mm}
\begin{tabular}{clcc}
\toprule
Scenario & Description & $\mathcal{B}_{95}^\mathrm{exp}(t\rightarrow u+g)$ & $\mathcal{B}_{95}^\mathrm{exp}(t\rightarrow c+g)$ \\ \midrule
(1)      & Data statistical only           & \num{1.1E-5} & \num{2.4E-5}~ \\
(2)      & Experimental uncertainties also & \num{3.1E-5} & \num{12E-5} \\
(3)      & All uncertainties except MC statistical & \num{3.9E-5} & \num{18E-5} \\
(4)      & All uncertainties               & \num{4.9E-5} & \num{20E-5} \\ \bottomrule
\end{tabular}
\end{center}
\end{table}
and clearly demonstrate how large the impact of systematic uncertainties is. Both the experimental and modelling uncertainties are relevant. MC statistical uncertainties increase the expected upper limits by approximately 20\% in the $ugt$ case and by about 10\% for the $cgt$ process.
 
\FloatBarrier
\section{Conclusions}
\label{sec:conclusions}
A search for the production of a single top quark via left-handed FCNC interactions of a top quark, a gluon and an up or charm quark was performed. The analysis used the full LHC Run~2 proton--proton collision data set recorded with the ATLAS detector at a centre-of-mass energy of 13$\,$\TeV, corresponding to an integrated luminosity of 139~fb$^{-1}$. Events with exactly one electron or muon, exactly one $b$-tagged jet and missing transverse momentum were selected, resembling the decay products of a single top quark. A dedicated high-purity working point was devised for the identification of $b$-jets, reducing the background of $W{+}c$-jets and $W$+light-flavour jets considerably. Neural networks were used to separate signal events from background events, and a binned maximum-likelihood fit to the neural-network discriminants was performed to search for a contribution from the $u+g\rightarrow t$ and $c+g\rightarrow t$ processes.
The observed distributions were found to be compatible with the background-only hypothesis and therefore upper limits on the production cross-sections times branching ratios were derived, leading to
\begin{eqnarray}
\sigma(ugt)\times\mathcal{B}(t\rightarrow Wb)\times\mathcal{B}(W\rightarrow \ell\nu) &<&3.0~\mathrm{pb} \ \ \ \ \mathrm{and} \nonumber \\
\sigma(cgt)\times\mathcal{B}(t\rightarrow Wb)\times\mathcal{B}(W\rightarrow \ell\nu) &<&4.7~\mathrm{pb}. \nonumber
\label{eq:xs_limits_conclusion}
\end{eqnarray}
The cross-section limits were interpreted in the framework of an effective field theory, yielding limits on the coefficients of the operators producing the FCNC processes under investigation: $|\CutuG|/\Lambda^2 < 0.057\,\TeV^{-2}$ and $|\CctuG|/\Lambda^2 < 0.14\,\TeV^{-2}$ at the 95\% confidence level. These limits are also expressed in terms of branching ratios of corresponding FCNC top-quark decays, resulting in
\[
\mathcal{B}(t\rightarrow u+g)<\num{0.61E-4} \ \ \ \ \mathrm{and} \ \ \ \ \mathcal{B}(t\rightarrow c+g)<\num{3.7E-4}.
\]
The new bounds improve on previous ATLAS results obtained at a centre-of-mass energy of $8\,\TeV$ by approximately a factor of two.

\section*{Acknowledgements}
 

We thank CERN for the very successful operation of the LHC, as well as the
support staff from our institutions without whom ATLAS could not be
operated efficiently.
 
We acknowledge the support of
ANPCyT, Argentina;
YerPhI, Armenia;
ARC, Australia;
BMWFW and FWF, Austria;
ANAS, Azerbaijan;
SSTC, Belarus;
CNPq and FAPESP, Brazil;
NSERC, NRC and CFI, Canada;
CERN;
ANID, Chile;
CAS, MOST and NSFC, China;
Minciencias, Colombia;
MSMT CR, MPO CR and VSC CR, Czech Republic;
DNRF and DNSRC, Denmark;
IN2P3-CNRS and CEA-DRF/IRFU, France;
SRNSFG, Georgia;
BMBF, HGF and MPG, Germany;
GSRI, Greece;
RGC and Hong Kong SAR, China;
ISF and Benoziyo Center, Israel;
INFN, Italy;
MEXT and JSPS, Japan;
CNRST, Morocco;
NWO, Netherlands;
RCN, Norway;
MEiN, Poland;
FCT, Portugal;
MNE/IFA, Romania;
JINR;
MES of Russia and NRC KI, Russian Federation;
MESTD, Serbia;
MSSR, Slovakia;
ARRS and MIZ\v{S}, Slovenia;
DSI/NRF, South Africa;
MICINN, Spain;
SRC and Wallenberg Foundation, Sweden;
SERI, SNSF and Cantons of Bern and Geneva, Switzerland;
MOST, Taiwan;
TAEK, Turkey;
STFC, United Kingdom;
DOE and NSF, United States of America.
In addition, individual groups and members have received support from
BCKDF, CANARIE, Compute Canada and CRC, Canada;
COST, ERC, ERDF, Horizon 2020 and Marie Sk{\l}odowska-Curie Actions, European Union;
Investissements d'Avenir Labex, Investissements d'Avenir Idex and ANR, France;
DFG and AvH Foundation, Germany;
Herakleitos, Thales and Aristeia programmes co-financed by EU-ESF and the Greek NSRF, Greece;
BSF-NSF and GIF, Israel;
Norwegian Financial Mechanism 2014-2021, Norway;
NCN and NAWA, Poland;
La Caixa Banking Foundation, CERCA Programme Generalitat de Catalunya and PROMETEO and GenT Programmes Generalitat Valenciana, Spain;
G\"{o}ran Gustafssons Stiftelse, Sweden;
The Royal Society and Leverhulme Trust, United Kingdom.
 
The crucial computing support from all WLCG partners is acknowledged gratefully, in particular from CERN, the ATLAS Tier-1 facilities at TRIUMF (Canada), NDGF (Denmark, Norway, Sweden), CC-IN2P3 (France), KIT/GridKA (Germany), INFN-CNAF (Italy), NL-T1 (Netherlands), PIC (Spain), ASGC (Taiwan), RAL (UK) and BNL (USA), the Tier-2 facilities worldwide and large non-WLCG resource providers. Major contributors of computing resources are listed in Ref.~\cite{ATL-SOFT-PUB-2021-003}.
 

\printbibliography

\clearpage 
 
\begin{flushleft}
\hypersetup{urlcolor=black}
{\Large The ATLAS Collaboration}

\bigskip

\AtlasOrcid[0000-0002-6665-4934]{G.~Aad}$^\textrm{\scriptsize 98}$,    
\AtlasOrcid[0000-0002-5888-2734]{B.~Abbott}$^\textrm{\scriptsize 124}$,    
\AtlasOrcid[0000-0002-7248-3203]{D.C.~Abbott}$^\textrm{\scriptsize 99}$,    
\AtlasOrcid[0000-0002-2788-3822]{A.~Abed~Abud}$^\textrm{\scriptsize 34}$,    
\AtlasOrcid[0000-0002-1002-1652]{K.~Abeling}$^\textrm{\scriptsize 51}$,    
\AtlasOrcid[0000-0002-2987-4006]{D.K.~Abhayasinghe}$^\textrm{\scriptsize 91}$,    
\AtlasOrcid[0000-0002-8496-9294]{S.H.~Abidi}$^\textrm{\scriptsize 27}$,    
\AtlasOrcid[0000-0002-9987-2292]{A.~Aboulhorma}$^\textrm{\scriptsize 33e}$,    
\AtlasOrcid[0000-0001-5329-6640]{H.~Abramowicz}$^\textrm{\scriptsize 157}$,    
\AtlasOrcid[0000-0002-1599-2896]{H.~Abreu}$^\textrm{\scriptsize 156}$
% ...
\bigskip
\\

% ...
$^{27}$Physics Department, Brookhaven National Laboratory, Upton NY; United States of America.\\
% ...
$^{33}$$^{(a)}$Facult\'e des Sciences Ain Chock, R\'eseau Universitaire de Physique des Hautes Energies - Universit\'e Hassan II, Casablanca;$^{(b)}$Facult\'{e} des Sciences, Universit\'{e} Ibn-Tofail, K\'{e}nitra;$^{(c)}$Facult\'e des Sciences Semlalia, Universit\'e Cadi Ayyad, LPHEA-Marrakech;$^{(d)}$LPMR, Facult\'e des Sciences, Universit\'e Mohamed Premier, Oujda;$^{(e)}$Facult\'e des sciences, Universit\'e Mohammed V, Rabat;$^{(f)}$Mohammed VI Polytechnic University, Ben Guerir; Morocco.\\
$^{34}$CERN, Geneva; Switzerland.\\
% ...
$^{51}$II. Physikalisches Institut, Georg-August-Universit\"{a}t G\"ottingen, G\"ottingen; Germany.\\
% ...
$^{91}$Department of Physics, Royal Holloway University of London, Egham; United Kingdom.\\
% ...
$^{98}$CPPM, Aix-Marseille Universit\'e, CNRS/IN2P3, Marseille; France.\\
$^{99}$Department of Physics, University of Massachusetts, Amherst MA; United States of America.\\
% ...
$^{124}$Homer L. Dodge Department of Physics and Astronomy, University of Oklahoma, Norman OK; United States of America.\\
% ...
$^{156}$Department of Physics, Technion, Israel Institute of Technology, Haifa; Israel.\\
$^{157}$Raymond and Beverly Sackler School of Physics and Astronomy, Tel Aviv University, Tel Aviv; Israel.\\
% ...
$^{a}$ Also at Borough of Manhattan Community College, City University of New York, New York NY; United States of America.\\
$^{b}$ Also at Bruno Kessler Foundation, Trento; Italy.\\
$^{c}$ Also at Center for High Energy Physics, Peking University; China.\\
$^{d}$ Also at Centro Studi e Ricerche Enrico Fermi; Italy.\\
$^{e}$ Also at CERN, Geneva; Switzerland.\\
$^{f}$ Also at D\'epartement de Physique Nucl\'eaire et Corpusculaire, Universit\'e de Gen\`eve, Gen\`eve; Switzerland.\\
% ...

\end{flushleft}


\end{document}